\documentstyle[10pt,aaspp4,flushrt]{article}

\begin{document}

%

\title{LIGHT PROPAGATION IN INHOMOGENEOUS UNIVERSES \\
       I: METHODOLOGY AND PRELIMINARY RESULTS}

\author{Premana Premadi,\altaffilmark{1,2} 
        Hugo Martel,\altaffilmark{3}
        and Richard Matzner\altaffilmark{1,2,4}}

\altaffiltext{1}{Center for Relativity, University of Texas, Austin, TX 78712}
\altaffiltext{2}{Department of Physics, University of Texas, Austin, TX 78712}
\altaffiltext{3}{Department of Astronomy, University of Texas, Austin, 
                 TX 78712}
\altaffiltext{4}{Orson Anderson Scholar, Los Alamos National Laboratory 
                 1996-97}

\begin{abstract}
We describe a numerical algorithm which simulates the propagation of light 
in inhomogeneous universes. This algorithm computes the trajectories of light
rays between the observer, located at redshift $z=0$, and distant sources
located at high redshift, using the multiple lens-plane method.
The deformation and deflection of light beams as they interact with each 
lens plane are computed using the filled-beam approximation.

We use a Particle-Particle/Particle-Mesh (P$^3$M) N-body numerical code
to simulate the formation of large scale structure in the universe.
We extend the length resolution of the simulations to sub-Megaparsec
scales by using a Monte-Carlo method for locating galaxies inside
the computational volume according to the underlying distribution
of background matter. The observed galaxy 2-point correlation function
is reproduced. This algorithm constitutes a major improvement over
previous methods, which either neglected the presence of large-scale
structure, neglected the presence of galaxies, neglected the
contribution of distant matter (matter located far from the beam),
or used the Zel'dovich approximation for simulating the formation of 
large-scale structure.
In addition, we take into account the observed morphology-density
relation when assigning morphological types to galaxies, something
that was ignored in all previous studies. 

To test this algorithm, we perform 1981 simulations, for three different
cosmological models: an Einstein-de~Sitter model with 
density parameter $\Omega_0=1$, an open model with $\Omega_0=0.2$, and a flat,
low density model with $\Omega_0=0.2$ and a cosmological 
constant $\lambda_0=0.8$. In all models, the
initial density fluctuations correspond to a Cold Dark Matter power spectrum
normalized to COBE.
In each simulation, we compute the shear and magnification resulting from
the presence of inhomogeneities. Our results are the
following: (1)~The magnification is totally dominated by the convergence,
with the shear contributing less than one part in $10^4$. 
(2)~Most of the cumulative
shear and magnification is contributed by matter located at intermediate 
redshifts $z=1-2$. (3)~The actual value of the redshift where the
largest contribution to shear and magnification occurs
depends on the
cosmological model. In particular, the lens planes contributing the most
are located at larger redshift for models with smaller $\Omega_0$.
(4)~The number of galaxies directly hit by the beam increases with redshift,
while the contribution of lens planes to the
shear and magnification decrease with 
increasing lens-plane redshift for $z>2$, indicating
that the bulk of the shear and magnification does not originate from direct 
hits, but rather from the tidal influence of nearby and
more distant galaxies, and background matter. 
(5)~The average contributions of background matter and nearby
galaxies to the shear is comparable for models with small $\Omega_0$.
For the Einstein-de~Sitter model, the contribution of the background
matter exceeds the one of nearby galaxies by nearly one order of magnitude.
\end{abstract}


\newpage
%

\section{INTRODUCTION}

Gravitational lensing systems have relatively simple geometry and most
of them are cosmological in nature because of their scale.
These two facts have stimulated cosmologists in the last few decades to
use gravitational lensing as a tool in physical cosmology.
There are four distinct applications of gravitational lenses to cosmology.
(a)~Observations of lensed sources provide information about the mass and
density structure of the lens (Zwicky 1937a,b). (b)~Lensed sources are 
amplified, and therefore easier to detect and resolve (Zwicky 1937a,b). 
(c)~Lensing of distant sources can be used to measure distances on 
cosmological scales (Klimov 1963; Liebes 1964; Refsdal 1964). (d)~Microlensing
can be used to study the stellar composition of the lens 
(Chang \& Refsdal 1979). For a review
of these various applications of gravitational lensing, we refer
the reader to Blandford \& Nayaran (1992), and references therein.
In this paper, we specifically focus on the problem of determining the 
nature and structure of the universe, using gravitational lensing of 
distant sources (quasars). 

The {\it nature} of the universe is described by Friedmann-\hbox{Lema\^\i tre} 
cosmological models. Such models describe idealized universes which are
structureless, and obey the weak cosmological principle (homogeneity
and isotropy). A matter-dominated
Friedmann-\hbox{Lema\^\i tre} model is characterized
by the present values of the density parameter $\Omega_0$ and the Hubble 
constant~$H_0$. All other cosmological parameters, such as the age of 
the universe $t_0$, the matter density $\bar\rho_0$, the deceleration
parameter $q_0$, or the curvature parameter $k$, can be expressed in terms of 
$\Omega_0$ and $H_0$. If the universe contains additional components, such
as radiation, cosmic strings, domain walls, or a nonzero cosmological 
constant, the model describing this universe has one additional parameter 
for each component, which measures the contribution of that component
to the energy density of the universe. 
Gravitational lensing of distant quasars can be used
to measure (or constrain) the value of these cosmological parameters, in
two different ways. First, the angular diameter distances between the source,
the lens, and the observer enter into the lens equation (Schneider,
Ehlers, \& Falco 1992, and references therein). 
Therefore, observables such as 
amplification, time delays, or image splitting will depend on these
distances. Since the relationship between the angular diameter distance
and the redshift is model-dependent
(Weinberg 1972; Fukugita et al. 1992), we can use observations of lensed
quasars to estimate $\Omega_0$ and $H_0$. Second, the probability that a
distant quasar will be lensed increases with the distance between that
quasar and the observer, since a larger distance implies a larger amount
of matter along the line of sight. Hence, by using the {\it fraction} of
lensed quasars, we can probe the distance-redshift relation and estimate
the cosmological parameters.

The {\it structure} of the universe represents the deviations from 
homogeneity and isotropy. These deviations are usually described in terms
of primordial fluctuations that grow with time as a result of 
gravitational instability, to eventually form the large-scale structures
of the universe, galaxies, clusters, and voids
(Peebles 1980). For most cosmological
models, the primordial density fluctuations originate from a Gaussian random
process, and therefore are characterized entirely by a density power
spectrum. There are numerous cosmological models describing the
formation of large-scale structures in the universe (Cold Dark Matter,
Hot Dark Matter, Mixed Dark Matter, $\ldots$), each model having
its own power spectrum. Since the formation and evolution of large-scale
structures are responsible for forming the lenses, observations of lensed
sources can be used to measure the amplitude and possibly the shape of the
power spectrum, and to ultimately determine the correct cosmological model
for large-scale structure formation.

To apply these methods, we need to study the propagation of light
in inhomogeneous universes described by particular
cosmological models. 
The most common approach consists of using numerical methods to simulate both 
the formation of large-scale structures in an expanding universe
and the propagation of photons through these structures.
With the existence of compact inhomogeneities in the universe, it is
reasonable to suspect that a light beam from a distant source undergoes
a series of perturbations while traveling to the observer. 
We can simulate the effect of these perturbations
by dividing the space between the source and the observer into 
redshifts intervals, and then projecting the matter inside each interval onto 
a plane normal to the line of sight, called a lens plane. In this
so-called {\it thin-lens approximation}, the deflection of light
resulting from each lens plane can be computed using geometrical optics.
We can follow the evolution of a light beam
propagating through the inhomogeneities,
adding successively the contributions of each lens plane to the deflection 
and deformation of the beam (Blandford \& Nayaran 1986; Blandford \& Kochanek
1987; Schneider \& Weiss 1988a, b; 
Jaroszy\'nski et al. 1990; Jaroszy\'nski 1991, 1992;
Babul \& Lee 1991; Bartelmann \& Schneider 1991; 
Wambsganss, Cen, \& Ostriker 1996; See also
Kochanek \& Apostolakis 1988; Paczy\'nski \& Wambsganss 1989). This
{\it multiple lens-plane method} is discussed in
detail in Schneider et al. (1992, Chap.~9).

To apply the multiple lens-plane method, 
we need to generate the surface density on each
lens plane. The simplest method consists of distributing
equal-mass objects (galaxies or dark halo) randomly in space,
Schneider \&~Weiss (1988b), Paczy\'nski \& Wambsganss (1989),
and Lee \& Paczy\'nski (1990)
have used this method to obtain statistics for shear and amplification
caused by gravitational lensing.
The obvious drawback of this approach is that it 
completely ignores the large-scale structure formation models that are
responsible for the formation of inhomogeneities, as well as the observed
properties of these inhomogeneities, such as the galaxy 2-point
correlation function or the morphology-density relation. 
These simulations are useful for studying the properties
of gravitational lenses, and their dependence upon cosmological parameters
such as $\Omega_0$ and $H_0$, but do not provide any information or
constraint on the large-scale structure formation scenario. Furthermore,
since the known properties of galaxy clustering in the universe are
ignored, these randomly-generated distributions of deflectors are 
unrealistic, and the relevance of the results is unclear.

Several authors have described analytical methods for
generating mass distributions that are consistent
with particular cosmological models. 
Babul \& Lee (1991) have developed an analytical model in which the effect
of large-scale structure enters the calculation of
light propagation through the density
auto-correlation function $\xi$. Since this function is the Fourier
transform of the density power spectrum, this method effectively
distinguishes among different models of structure formation. 
Bartelmann \& Schneider (1991) use the semi-analytical model of
Buchert (1989) to generate large-scale structure for
an Einstein-de~Sitter model with a flat perturbation spectrum.
Jaroszy\'nski (1991, 1992) generates initial density fluctuations
that reproduces a particular power spectrum, and then
uses the Zel'dovich approximation to simulate the 
evolution of these density fluctuations. He then locates galaxies
in each lens plane using an empirical method based on the local 
matter density. 
By combining the Schechter luminosity function
with a Monte Carlo method, he generates a luminosity for each
galaxy, chooses a morphological type at random,
and then model that galaxy using a non-singular isothermal profile
whose parameters are related to the luminosity and morphological type
of the galaxy. 
This constitutes a major improvement over previous work, in
three different ways. (a)~The method takes into account the fact that the
large-scale structures in the universe do originate from the
growth of primordial fluctuations, and allows for experimentations
with different power spectra. (b)~The galaxies have a spectrum of
luminosity and masses that reproduce observations.
(c)~The method takes into account the existence of
various galaxy morphological types (ellipticals, S0's, and spirals), by 
ascribing different surface density profiles to galaxies of different types.

There are still several weaknesses in the approach used by Jaroszy\'nski.
First, the Zel'dovich approximation is based on linear perturbation 
theory, and therefore underestimates the growth of large-scale structures
in overdense regions. This can be a serious problem since these
overdense regions are the one most likely to affect significantly the
evolution of the beam. 
Second, while this method acknowledges the existence
of various morphological types, the morphological type of each galaxy is
chosen randomly. This ignores the existence of the {\it Morphology-Density
Relation} (Dressler 1980; Postman \& Geller 1984), which relates the likelihood
of any galaxy to have a particular morphological type to the richness of 
the environment in which that galaxy is located.
 
In an earlier paper (Jaroszy\'nski et al. 1990),
a Particle-Mesh [PM] code was used for generating the large-scale
structure of the universe. This approach constitutes a significant 
improvement over using the Zel'dovich approximation.
However, that paper did not include a treatment 
of the galaxies similar to the one in Jaroszy\'nski (1991, 1992).
Blandford et al. (1991) also used a PM code for generating the large-scale
structure, and also considered the alternative approach of representing
galaxies by randomly distributed isothermal spheres. However, they did
not take the additional step of combining the results of the PM
simulations with the density profiles of galaxies, as we shall do
in this paper.

There is also a potential problem with some of the methods described above.
In several cases, only the matter located near the beam is included in the
calculation of the deflection and deformation of the beam. The influence
of distant matter is neglected. Neglecting the contribution of distant matter
is probably correct if the galaxy distributions are generated randomly,
but, as we said, these galaxy distribution are unrealistic to start with.
The analytical methods and PM simulations described above
allow the formation of large-scale structures
such as clusters of galaxies. As Blandford et al. (1991) showed,
the effect of distant clusters on the evolution of the beam can be important
(this is supported by the analytical work of Kaiser [1992]).

In this paper, we present a new method which addresses all these 
various concerns. In designing this method, our
main goal was to generate matter distributions that take into
account all the known constraints imposed by large-scale structure
formations models and by observations of the actual distribution,
morphological types, and structure of galaxies in the universe. 
To achieve this goal, we use a state-of-the-art 
Particle-Particle/Particle-Mesh
($ {\rm P}^{3}{\rm M}$) code (Hockney \& Eastwood 1981) to
simulate the formation and evolution of large-scale structure in 
the universe. For all simulations presented in this paper, we
use a Cold Dark Matter density power spectrum normalized to COBE,
but the method can be used with any power spectrum and any normalization.
Using the particle distributions generated by the P$^3$M code,
we locate the galaxies (in the densest regions),
using a Monte Carlo method that reproduces the observed galaxy 2-point 
correlation function fairly well (within the limitation of 
CDM model). Then, instead of randomly choosing the morphological type of each 
galaxy, a method
which ignores the existence of morphological segregation, we determine the
morphological type of each galaxy according to the local environment, using
the observed morphology-density relation (Martel, Premadi, \&~Matzner 1997a,
hereafter MPM). Each galaxy is given a surface density profile which is
chosen according to the galaxy luminosity and morphological type,
as in Jaroszy\'nski (1992). By combining the distribution of background matter
simulated by the P$^3$M algorithm with the distribution and surface
densities of galaxies, we are effectively describing the surface
density of the lens planes over 9 orders of magnitude in length, from the
size of the largest superclusters and voids, $\sim100\,\rm Mpc$, down
to the core radii of the smallest galaxies, $\sim0.1\,\rm pc$.

This approach for generating the surface density on the lens planes, the
key part of any multiple lens-plane algorithm, differs 
significantly from all the ones
that have been published previously. In their early work, 
Schneider \& Weiss (1988a, b) distributed clumps of equal masses randomly
on the lens planes, thus ignoring both the existence of large-scale 
structure and the mass spectrum and structure of galaxies. Jaroszy\'nski
et al. (1990) and Wambsganss et al. (1996) simulated the formation of
large-scale structure, but did not take galaxies into account. Blandford
et al. (1991) performed N-body simulation of large-scale structure formation,
and also computed
the deflection of light by randomly distributed galaxies. However,
they considered these two approaches 
separately, and did not choose 
the location of the galaxies according to the results
of the N-body simulations, as we do in this paper. The only algorithm
which combines large-scale structure formation with galaxies is
the one described by Jaroszy\'nski (1991, 1992). However, the evolution of
the large-scale structure in that algorithm was simulated using the Zel'dovich
approximation instead of a N-body code. Furthermore, the contribution of 
distant matter to the evolution of the beam was ignored.

Our algorithm is also the first that takes the morphology-density relation
into account. The reason is clear: none of the previous algorithms 
could have done it, either because galaxies
were ignored, or the process of cluster formation was either ignored
or approximated, making the morphology-density relation unapplicable.
The combination of fully nonlinear large-scale structure formation,
galaxy distributions that reproduce the observed 
2-point correlation function, 
morphological type distributions that reproduce the observed morphology-density
relation, and galaxy surface density profiles, gives to the matter distribution
in our algorithm a level of realism that was not present in any
of the previous studies.

Of course, the algorithms used by previous authors can be perfectly
adequate, depending on the particular problem that is being studied,
and have produced very interesting results. We feel, however, that for the
purpose of determining the correct cosmological model of structure formation
in the universe, and the value of the cosmological parameters, it is
critical to generate matter distribution that are as realistic
as possible, over the largest possible range of length scales. This was
our goal in designing this algorithm.

We briefly review the theory of gravitational lensing in \S2, 
mentioning only the aspects which are directly relevant to this paper. 
In \S3
we describe the method for generating
the large-scale mass distribution of the universe. The simulation
of the light propagation and the resulting statistics are
described in \S4. Summary and conclusion are presented in \S5.
 
\section{BASIC THEORY OF GRAVITATIONAL LENSING}

\def \etabf   {\eta}
\def \thetabf {\theta}
\def \deltabf {\delta}
\def \alphabf {\alpha}
\def \betabf  {\beta}
\def \xibf    {\xi}

\subsection{The Geometry}

With the existence of compact inhomogeneities in our universe, it is
reasonable to suspect that a light bundle from a distant source undergoes
a series of perturbations due to those inhomogeneities while 
traveling to the observer. We attempt to use a series of gravitational lensings
to approximate this phenomena. First, we idealize the inhomogeneities
as being distributed on thin sheets,
called lens planes, which are arranged perpendicular
to the line of sight. We assume that lensing only takes place on
each of those planes. This way we can analyze
the lensing properties of each plane separately, and let the light beam
carries the effect of lensing while propagating from one plane to the next.
This is known as the multiple lens-plane method
(Schneider et al. 1992).

Consider $N$ lens planes located at redshifts $z_{i}$,
with $i=1,N$, and ordered such that $z_{i}<z_{j}$ for $i<j$. 
Figure~1 shows an example with $N=2$. All angles are
greatly exaggerated. Each lens plane is characterized by its
respective surface mass density $\sigma_{i}(\xibf_{i})$,
where $\xibf_{i}$ is the impact vector of the ray on the 
$i$-th lens plane. Let $\hat{\alphabf}_i(\xibf_{i})$ 
denote the
deflection angle the light ray experiences on the $i$-th plane at a
position $\xibf_{i}$. {From} this geometry, we can derive the 
{\it lens equation},
\begin{equation}
\etabf = {D_S\over D_1}\xibf_1 - \sum_{i=1}^{N} D_{iS}
            \hat{\alphabf}_i(\xibf_{i}) \,,
\end{equation}

\noindent where $\etabf$ is source position vector (on the source plane),
$\xibf_{i}$ is the impact vector on the $i$-th plane, 
$D_j$ is the angular diameter distance between
the $j$-th plane and the observer, and $D_{ij}$ is the angular diameter
distance between the $i$-th and $j$-th planes, with $S\equiv N+1$ 
identifying the source plane.
Knowing the impact vector $\xi_1$ on the image plane,
the impact vector on subsequent planes can be 
obtained recursively using
\begin{equation}
\xibf_{j} = {D_j\over D_1}\xibf_1 - \sum_{i=1}^{j-1} D_{ij} 
            \hat{\alphabf}_i(\xibf_{i}) \,.
\end{equation}

\noindent
The deflection angle is related to the surface density by
\begin{equation}
\hat{\alphabf}_i(\xibf)={4G\over c^2}\int\!\!\!\int\sigma_i(\xibf')
               {\xibf_i-\xibf'\over|\xibf_i-\xibf'|^2}d^2\xi'\,,
\end{equation}

\noindent
where $G$ is the gravitational constant, $c$ is the speed of light,
and the integral extends over the lens plane. We can rewrite this expression
conveniently as
\begin{eqnarray}
\hat{\alphabf}_i(\xibf_i)&=&\nabla\hat\psi_i(\xibf_i)\,,\\
\hat\psi_i(\xibf_i)&=&{4G\over c^2}\int\!\!\!\int
         \sigma_{i}(\xibf')\ln |\xibf_i-\xibf'|d^2\xi'\,.
\end{eqnarray}

\noindent It is useful to rewrite these equations
in a dimensionless form. We define
for each lens plane a critical surface density as
\begin{equation}
\sigma_{i,\rm cr}={c^2D_S\over4\pi GD_iD_{iS}}\,,
\end{equation}

\noindent
and introduce the following dimensionless quantities,
\begin{eqnarray}
{\bf x}_{i}&=&{\xibf_i\over D_i}\,,\qquad 1 \leq i \leq N+1 \,;\\
\kappa_{i}({\bf x}_i)
&=&{\sigma_i\over\sigma_{i,\rm cr}}\,,\qquad 1 \leq i \leq N \,.
\end{eqnarray}

\noindent Equations (2), (4), and (5) reduce to
\begin{eqnarray}
{\bf x}_{j} &=& {\bf x}_{1} - \sum_{i=1}^{j-1} \beta_{ij}
                \alphabf_{i}({\bf x}_{i})\,,\\
\alphabf_{i}({\bf x}_i) &=& \nabla \psi_{i}({\bf x}_i)\,,\\
\psi_{i}({\bf x}_{i})&=& {1\over\pi}\int\!\!\!\int
         \kappa_{i}({\bf x}') \; \ln |{\bf x}_i-{\bf x}'|d^2x'\,,
\end{eqnarray}

\noindent where
\begin{equation}
\beta_{ij}={D_{ij}D_S\over D_jD_{iS}}\,,
\end{equation}

\noindent 
and the gradient is now taken relative to ${\bf x}_i$. 
By using the identity $\nabla^2\ln|{\bf x}_i|=2\pi\delta^2({\bf x}_i)$
(where $\delta^2$ is the two-dimensional delta function),
we can invert equation~(11), and get
\begin{equation}
\nabla^{2}\psi_i=2\kappa_i\,.
\end{equation}

\noindent
To compute the scaled position ${\bf y}\equiv{\bf x}_S$ of the source
on the source plane, we simply set $j=N+1$. Equation~(12) gives
$\beta_{iS}=1$, and equation~(9) becomes
\begin{equation}
{\bf y} \equiv {\bf x}_{N+1} = {\bf x}_{1} - \sum_{i=1}^{N} 
                                \alphabf_{i}({\bf x}_{i})\,.
\end{equation}

\noindent
This ray-tracing equation is a mapping from the image plane 
($i=1$) onto the source plane ($i=N+1$).

\subsection{Angular-Diameter Distances}

Since we are using the filled-beam approximation, the relevant distances
to use in the lens equations are the angular diameter distances in an 
homogeneous Friedmann Universe
(Schneider \& Weiss 1988a). For the cosmological models considered in
this paper, all the appropriate distance formulae are given in Fukugita et 
al. (1992).
For the Einstein-de~Sitter model, the angular diameter distance $D$
between redshifts $z_i$ and $z_j$ is
\begin{equation}
D(z_i,z_j)={2R_0\over1+z_j}\Big[(1+z_i)^{-1/2}-(1+z_j)^{-1/2}\Big]\,,
\end{equation}

\noindent where $R_0=c/H_0$ is the Hubble radius. For the open model with
$\Omega_0<1$, the distance is
\begin{equation}
D(z_i,z_j)={2R_0\over\Omega_0^2(1+z_i)(1+z_j)^2}\Big[
(2-\Omega_0+\Omega_0z_j)(1+\Omega_0z_i)^{1/2}
-(2-\Omega_0+\Omega_0z_i)(1+\Omega_0z_j)^{1/2}\Big]\,.
\end{equation}

\noindent Finally, for the flat model with nonzero cosmological constant,
the distance is
\begin{equation}
D(z_i,z_j)={R_0\over1+z_j}\int_{z_i}^{z_j}dz\Big[\Omega_0(1+z)^3
+(1-\Omega_0)\Big]^{-1/2}\,.
\end{equation}

\noindent Of course, equation~(15) is a special case of both 
equations~(16) and~(17). In all cases, we are assuming $z_j>z_i$, which 
gives $D>0$. Figure~2 shows the angular diameter distances 
$D_S\equiv D(0,z_s)$ for all three cosmological models considered in this 
paper.

\subsection{The Magnification Matrix}

The effect of each lens plane on the evolution of the beam 
is described by the following Jacobian matrix,
\begin{equation}
{\bf A}_i({\bf x}_i)={\partial {\bf x}_{i+1}\over\partial {\bf x}_i}
=\left(\matrix{1-\psi_{i,11} &  -\psi_{i,12} \cr
                -\psi_{i,21} & 1-\psi_{i,22} \cr}\right)\,,
\end{equation}

\noindent
where the commas denote differentiation
with respect to the components of ${\bf x}_i$.
Since $\psi_{i,12}=\psi_{i,21}$, and equation~(13) gives 
$\psi_{i,11}+\psi_{i,22}=2\kappa_i$, we can rewrite equation~(18) as
\begin{equation}
{\bf A}_i=\left(\matrix{1-\kappa_i-S_{11} & -S_{12}\cr
-S_{12} & 1-\kappa_i+S_{11} \cr }\right)\,,
\end{equation}

\noindent where
\begin{eqnarray}
S_{11}&=&{1\over2}(\psi_{i,11}-\psi_{i,22})\,,\\
S_{12}&=&\psi_{i,12}=\psi_{i,21}\,.
\end{eqnarray}

\noindent We now define
\begin{equation}
S_i=(S_{11}^2+S_{12}^2)^{1/2}\,.
\end{equation}

\noindent The determinant and trace of 
${\bf A}_i$ can be
expressed entirely in terms of $\kappa_i$ and $S_i$, as follows:
\begin{eqnarray}
\det\,{\bf A}_i&=&(1-\kappa_i)^{2}-S_i^{2}\,,\\
{\rm tr}\,{\bf A}_i&=&2(1-\kappa_i)\,.
\end{eqnarray}

\noindent The quantities $\mu_i\equiv1/(\det{\bf A}_i)$, $1-\kappa_i$, and 
$S_i$ are called {\it magnification}, {\it convergence} (or Ricci focusing),
and {\it shear}, respectively.

To compute the cumulative effect of all the lens planes, we consider
the Jacobian matrix of the mapping given by equation~(14),
\begin{equation}
{\bf B}({\bf x})={\partial{\bf y}\over\partial{\bf x}_1}
={\bf I}-\sum_{i=1}^N{\partial\alphabf_i\over\partial{\bf x}_1}
={\bf I}-\sum_{i=1}^N{\bf U}_i{\bf B}_i\,,
\end{equation}

\noindent where $\bf I$ is the $2\times2$ identity matrix,
and ${\bf U}_i$ and ${\bf B}_i$ are defined by
\begin{eqnarray}
{\bf U}_i&=&{\partial\alphabf_i\over\partial{\bf x}_i}\,,\\
{\bf B}_i&=&{\partial{\bf x}_i\over\partial{\bf x}_1}\,.
\end{eqnarray}

\noindent After substituting equation~(10) into equation~(26), we
get 
\begin{equation}
{\bf U}_i={\bf I}-{\bf A}_i=
\biggl(\matrix{ \psi_{i,11} & \psi_{i,12} \cr
                \psi_{i,21} & \psi_{i,22} \cr }\biggr)
\,,
\end{equation}

\noindent where ${\bf A}_i$ is given by equation~(18).
Hence, ${\bf U}_i$ describes
the effect the $i$-th plane would have on the beam if all
the other planes were absent, and equation~(25) simply combines the
effect of all the planes. To compute the matrices ${\bf B}_i$,
we differentiate equation~(9), and get
\begin{equation}
{\bf B}_j={\bf I}-\sum_{i=1}^{j-1}\beta_{ij}{\bf U}_i{\bf B}_i\,.
\end{equation}

\noindent Since ${\bf B}_1={\bf I}$, we can use equation~(29) to compute
all matricies ${\bf B}_i$ by recurrence.

The image of a small circular source\footnote{We consider a source to be 
``small'' if the matrix ${\bf B}$ is essentially constant across the area of 
the source.} is an ellipse with semi-axes $r/\lambda_1$ and $r/\lambda_2$,
where $r$ is the radius of the image in the absence of lensing.
It can be shown that
\begin{eqnarray}
\lambda_1\lambda_2&=&\det{\bf B}\,, \\
\lambda_1^2+\lambda_2^2&=&{\rm tr}({\bf BB}^t)\,.
\end{eqnarray}

\noindent We can solve these equations for $\lambda_1$ and $\lambda_2$.
Assuming $\lambda_1\geq\lambda_2$, we can then 
compute the aspect ratio of the image. After some algebra, we get
\begin{equation}
{\lambda_1\over\lambda_2}
={{\rm tr}({\bf BB}^t)+\left\{[{\rm tr}({\bf BB}^t)]^2
-4(\det{\bf B})^2\right\}^{1/2}\over2\det{\bf B}}\,.
\end{equation}

\noindent The magnification is given
by 
\begin{equation}
\mu=(\lambda_1\lambda_2)^{-1}=(\det{\bf B})^{-1}\,.
\end{equation}

\section{THE NUMERICAL ALGORITHM}

\subsection{Overview}

Light rays coming from distant sources are propagating through the 
universe while the large-scale structure in the universe is
forming and evolving. Ultimately, it would be desirable to simulate
the evolution of large-scale structure and the light propagation
simultaneously. However, such approach would be rather difficult, and is
beyond the reach of present computer capabilities. The great
advantage of using the multiple lens-plane method is that it allows
us to consider the large-scale structure evolution problem and
the light propagation problem separately, thus effectively
breaking up the problem into two steps. The first step consists
of generating the large-scale structure and galaxy distribution in 
the universe at various redshifts, and projecting these distributions
onto lens planes, normal to the optical axis.
Then, once these lens planes are generated, we can compute numerically
the trajectory of light rays through them, using the formalism
described in \S2. Clearly, many different experiments can be
conducted using the same set of lens planes, simply by varying the 
shape, size, and number of rays in the beam, or the location of the beam on
the planes. 

In \S\S3.2 and~3.3, we describe the method we use for generating galaxy
distributions. Several aspects of this method were 
previously discussed in detail in MPM, so we only give a brief summary. 
The ray-shooting method is described in \S3.4.

\subsection{Large-Scale Structure Formation}

\subsubsection{The $\rm P^3M$ Algorithm}

All N-body simulations presented in this paper
are done using the P$^3$M algorithm (Hockney \&~Eastwood 1981). 
The calculations evolve a system of 
gravitationally interacting particles in a cubic volume with
triply periodic boundary conditions, comoving with Hubble flow.
The forces on particles are computed by solving Poisson equation on a 
cubic grid using a Fast Fourier Transform method.
The resulting force field represents the Newtonian interaction
between particles down to a separation of a few mesh spacings. At shorter
distances the computed force is significantly smaller than the
physical force. To increase
the dynamical range of the code, the force at short distance 
is corrected by direct 
summation over pairs of particles separated by less than some 
cutoff distance~$r_e$. With the addition of this so-called
{\it short-range correction}, the code accurately reproduces the Newtonian
interaction down to the softening length~$\eta$.
In all calculations, $\eta$ and $r_e$ were set equal to
0.3 and 2.7 mesh spacing, respectively,
with $64^3$ particles and a $128^3$ grid. With these particular values, 
the code has a dynamical range of three orders of magnitude in length.
The system is evolved forward in time using a 
second order Runge-Kutta time-integration scheme with a variable time step.

\subsubsection{Redshift of the Lens Planes}

To implement the multiple lens-plane method, we divide the space between $z=0$ 
and $z=5$ into a chain of cubic boxes of equal comoving size~$L_{\rm box}$. 
We first need to determine the
redshifts of the interfaces between these cubic boxes. Let us assume that
the photons that are reaching the observer at present entered a particular
box at time $t'$, redshift $z'$ and exited that box at time $t$,
redshift $z$. The redshifts $z'$ and $z$ are related by
\begin{equation}
L_{\rm box}=\int_{t'}^{t}\big[1+z(t)\big]c\,dt\,,
\end{equation}

\noindent where $c$ is the speed of light.
Using this equation, with the appropriate relation for $z(t)$,
we can find the redshifts of the interfaces. The front
side of the box closest the the observer is, by definition, at $z=0$. Plugging
this value into equation~(34) gives us the redshift $z'$ of
the back side of the box, which is also the redshift $z$ of the front side
of the next box. Then, by using equation~(34) recursively, 
we can compute the redshifts of all the interfaces. The derivation
of the recurrence relations for the Einstein-de~Sitter model 
($\Omega_0=1$, $\lambda_0=0$), open models ($\Omega_0<1$, $\lambda_0=0$),
and flat models with a nonzero cosmological constant 
($\Omega_0+\lambda_0=1$) are presented in detail in Premadi (1996).

Next, we need to determine the matter distribution inside each box. However,
during the time photons propagate across a particular box, 
the matter distribution inside that box evolves. In the thin-lens
approximation, we need to choose for each box a ``snapshot redshift''
$z_{\rm snap}$ between the redshift $z'$ when the photons enter the box and
the redshift $z$ when the photons exit the box, generate the matter
distribution at that redshift, and make the approximation that this 
distribution is valid at all redshifts between $z'$ and $z$. Then, we need
to choose a ``projection redshift'' $z_{\rm proj}$, also between $z'$ and $z$,
which is the redshift of the plane onto which we project the three-dimensional
distribution of galaxies.

We decided to set the projection redshift $z_{\rm proj}$ equal to the
snapshot redshift $z_{\rm snap}$, as every other author does. 
Schneider \&~Weiss (1988b, eq.~[8]) choose for the snapshot redshift
$z_{\rm snap}$ the arithmetic mean $(z'+z)/2$.
We decided to improve on this, by determining the snapshot
redshift as follows: Since the deflection angle varies linearly
with the surface density of the lens
plane, we choose $z_{\rm snap}$ to be the redshift
at which the density contrast $\delta$ is equal, at each point, to the 
time-averaged
value $\bar\delta$ of the density contrast at that point between $z'$ and $z$
(this only makes sense in the context of linear perturbation theory,
where the density contrast at any given point evolves independently of the
density contrast at other points). For the Einstein-de~Sitter model,
the linear
density contrast $\delta=K(t/t_0)^{2/3}$, where $t_0$ is the present time
and $K$ is a constant. The time-averaged 
linear density contrast between two epochs
$t'$ and $t$ is then given by
\begin{equation}
\bar{\delta}\equiv{1\over t-t'}\int_{t'}^t\delta(t)dt=
{3K\over5}\left[{(t/t_{0})^{5/3}-(t'/t_{0})^{5/3}\over
     (t/t_{0})-(t'/t_{0})}\right]\,.
\end{equation}

\noindent We set $\bar\delta=K(\bar t/t_0)^{2/3}=(1+z_{\rm snap})^{-1}$, and 
solve for $z_{\rm snap}$ as a function of $z'$ and $z$. We get
\begin{equation}
z_{\rm snap}={5\over3}\left[{(1+z)^{-3/2}-(1+z')^{-3/2}\over
  (1+z)^{-5/2}-(1+z')^{-5/2}}\right]-1\,.
\end{equation}

\noindent Computing $z_{\rm snap}$ for the other models is a significantly 
more complicated procedure, and constitutes an overkill. Equation~(36) 
is valid at high redshift, where all
models resemble the Einstein-de~Sitter model.
At low redshift, linear theory is inaccurate whether we use the correct model
or not. However, equation~(36) reduces to the
Schneider \&~Weiss formula $z_{\rm snap}=(z'+z)/2$ in
the low redshift limit.
Hence, it is correct to use equation~(36) for all cosmological models,
and it constitutes an improvement over the formula used by Schneider 
\& Weiss~(1988b)

\subsection{The Galaxies Distributions}

\subsubsection{The Galaxy Locations}

The method we use for computing
the galaxy locations was described in great detail in MPM. 
In this subsection, we give a brief summary
of the method. It consists of three parts. First, we determine the locations
of the galaxies at $z=0$. Second, we ascribe to each galaxy a morphological
type (E, S0, or Spiral). Finally, we trace the galaxies back in time to 
determine their locations on each lens plane.

We consider the large-scale structure at present ($z=0$) resulting from
the P$^3$M simulations, and design an empirical Monte-Carlo method
for locating galaxies in the computational volume, based on the
constraints that (1) galaxies should be predominantly located in the
densest regions, and (2) the resulting distribution of galaxies
should resemble the observed distribution on the sky.
Our method is the following: we divide the present
computational volume into $128^3$ cubic cells of size $1\,{\rm Mpc}^3$, and
compute the matter density $\rho$ at the center of each 
cell, using the same mass assignment scheme
as in the P$^3$M code. We then choose a particular density
threshold $\rho_{\rm t}$. We locate $N$ galaxies in each cell, where
$N$ is given by
\begin{equation}
N={\rm int}\biggl({\rho\over\rho_{\rm t}}\biggr)\,.
\end{equation}

\noindent The actual location of each galaxy is chosen to be
the center of the cell, plus a random offset of order of the cell size.
This eliminates any spurious effect introduced by the use of a grid.
We then experiment with various values of the density
threshold $\rho_{\rm t}$ until the total number of galaxies comes out to be
of order 40000. This gives a number density of 
$\sim0.02\,{\rm galaxies}/{\rm Mpc}^3$. This method bears some
similarities with the one used by Jaroszy\'nski (1991, 1992).
Tests showed that the observed galaxy 2-point correlation function
is fairly well reproduced (MPM).

There is a well-known observed relationship between the distribution
of morphological types and the surface density of galaxies (Dressler 1980;
Postman \& Geller 1984). Regions of the sky with high concentration of
galaxies contain on average more elliptical and S0's and less spiral than
regions with lower concentration of galaxies. By combining
this relation with a Monte-Carlo method, we can ascribe a morphological 
type to each galaxy, as follows. 
We first compute the volume number density of galaxies $\rho_{\rm gal}$
around each galaxy, using 
\begin{equation}
\rho_{\rm gal}={n+1\over4\pi d_n^3/3}\,,
\end{equation}

\noindent where $n$ is a positive integer, and $r_n$ is the distance of the
$n^{\rm th}$ nearest neighboring galaxy. In all cases, we choose $n=12$.
Once the densities are computed, we compute the fractions 
$f_{\rm Sp}(\rho_{\rm gal})$, $f_{\rm S0}(\rho_{\rm gal})$,
and $f_{\rm Ell}(\rho_{\rm gal})$ of spirals, S0's, and ellipticals,
respectively,
from the morphology-density relation. We then ascribe a 
morphological type to
each galaxy by generating a random number $x$ between 0 and 1 (with
uniform probability). The galaxy is identified as a spiral if
$x<f_{\rm Sp}$, a S0 if $f_{\rm Sp}<x<f_{\rm Sp}+f_{\rm S0}$, and
an elliptical if $x>f_{\rm Sp}+f_{\rm S0}$. 

The P$^3$M algorithm provides us with the distributions of particles
at various intermediate redshifts between the initial redshift
and the present, and, in particular, at all
snapshots redshifts $z_{i,{\rm snap}}$, $i=1,\ldots,N$. 
By combining these particle distributions with our
simulated galaxy distributions at present, we can trace galaxies back in time
and reconstruct their trajectories. To do this, we simply find the nearest 
particle $p_k^{(1)}$ of each galaxy $g_k$ at present
(where the subscript $k$ identifies the galaxy). Then we ``tie''
the galaxy $g_k$ to that nearest particle. The location of the galaxy
$g_k$ at any redshift $z$ is then given by:
\begin{equation}
{\bf r}(g_k,z)={\bf r}\Big[p_k^{(1)},z\Big]+{\bf r}'\,,
\end{equation}

\noindent where ${\bf r}'$ is a small random offset, which we introduce to
avoid the unfortunate situation of having two galaxies located at the
top of each other because they happen to by tied to the same
particle. This allows us to construct galaxy distributions at any redshift.

\subsubsection{The Galaxy Parameters}

To determine the physical parameters of each galaxy, we
start by assuming that the present galaxy luminosities follow
the Schechter luminosity function,
\begin{equation}
n(L)dL={n_*\over L_*}\left({L\over L_*}\right)^{\alpha}e^{-L/L_*}dL\,,
\end{equation}

\noindent
where $n(L)$ is the number density of galaxies per unit luminosity. The 
parameters $n_*$, $L_*$, and $\alpha$ are obtained from observation as
follows:
 $\alpha=-1.10$, $n_*=0.0156h^{3}{\rm Mpc}^{-3}$, and $L_{B*}=1.3\times
10^{10}h^{-2}L_{\odot}$, where $L_B$ is the luminosity
in the $B$ band (Efstathiou, Ellis, \& Peterson 1988, hereafter EEP). 
There is a fourth parameter, 
the luminosity $L_{\min}$ of the faintest galaxies,
which must be introduced to prevent the total number of galaxies
from diverging. We now make 
the assumption that the numerical values of $\alpha$ and $L_*$ given in
EEP are quite reliable, but the numerical values of
$n_*$ and $L_{\min}$ might be less reliable, because of the difficulty
of detecting galaxies at the low-luminosity end. Instead, we shall solve for
the values of $n_*$ and $L_{\min}$. This requires two constraints. We impose
that the mean density $n_0$ of galaxies matches the value of 
$0.02\,\rm Mpc^{-3}$ that we assume in our simulations, and that the mean 
luminosity density matches the value 
$j_0=1.93\times10^8hL_\odot\,\rm Mpc^{-3}$ given in EEP.

Equation~(40) allows us to directly compute the present number density
$n_0$, and luminosity density $j_{0}$, 
\begin{eqnarray}
n_0&=&n_*\int_{x_{\min}}^{\infty} x^{\alpha} e^{-\alpha}dx\,,\\
j_0&=&n_*L_*\int_{x_{\min}}^{\infty}x^{\alpha+1}e^{-x}dx
     =L_*\Big[n_*\; x_{\min}^{\alpha+1}\;e^{-x_{\min}} 
            + n_0(\alpha+1)\Big]\,,
\end{eqnarray}

\noindent
where $x\equiv L/L_*$ and $x_{\min}\equiv L_{\min}/L_*$.
The last equality in equation~(42) was obtained by integrating by part,
and then substituting in equation~(41).
We now substitute the numerical values of
$L_*$, $j_0$, and $\alpha$ (with $h=0.5$), and get,
\begin{eqnarray}
n_*\int\limits_{x_{\min}}^\infty x^{-1.1}e^{-x}dx&=&0.02\,{\rm Mpc}^{-3}\,,\\
n_*x_{\min}^{-0.1}e^{-x_{\min}}&=&3.86\times10^{-3}{\rm Mpc}^{-3}\,.
\end{eqnarray}

\noindent This system of equations can be solved 
numerically for $n_\ast$ and $x_{\min}$. The solution is
\begin{eqnarray}
n_\ast&=&0.00174\,{\rm Mpc}^{-3}\,,\\
x_{\min}&=&3.50095\times10^{-4}\,.
\end{eqnarray}

\noindent
Since, for all calculations presented in this paper,
we assume a Hubble constant of $50\,\rm km\,s^{-1}Mpc^{-1}$, equation~(45)
can be rewritten as $n_*=0.0139h^3\rm Mpc^{-3}$, which is, within
error bars, consistent with the value given in EEP.

We adopt the galaxy models described in Jaroszy\'nski (1991,1992). 
The projected surface density of each galaxy is given by
\begin{equation}
\sigma(r) = \cases{\displaystyle
            {v^2\over4G(r^2+r_c^2)^{1/2}}\,,  & $r<r_{\max}$\,; \cr
             0\,,                             & $r>r_{\max}$\,; \cr}
\end{equation}

\noindent where $r$ is the projected distance from the center.
The parameters $r_c$, $r_{\max}$, and $v$ are the core radius, maximum radius,
and velocity dispersion, respectively, and are given by
\begin{equation}
r_{c}=\cases{
         100 h^{-1} \left(\displaystyle{L\over L_*}\right){\rm pc}\,, & 
             (Ellipticals and S0's)\,; \cr
         1 h^{-1} \left(\displaystyle{L\over L_*}\right){\rm kpc}\,,  &
             (Spirals)\,; \cr}
\end{equation}

\begin{equation}
r_{\max}=30 h^{-1} \left(\displaystyle{L\over L_*}\right)^{1/2}{\rm kpc}\,;
\end{equation}

\begin{equation}
v=\cases{
390\;{\rm km\,s^{-1}} \left(\displaystyle{L\over L_{*}}\right)^{1/4}\,,
& (Ellipticals)\,; \cr
357\;{\rm km\,s^{-1}} \left(\displaystyle{L\over L_{*}}\right)^{1/4}\,,
& (SO's)\,; \cr
190\;{\rm km\,s^{-1}} \left(\displaystyle{L\over L_{*}}\right)^{0.381}\,,
& (Spirals)\,. \cr}
\end{equation}

We use a Monte-Carlo method to generate for each galaxy a luminosity
$L\geq L_{\min}$, with a probability $P(L)$ proportional to
$n(L)$. This ensures that the ensemble of $\sim40000$ galaxies
in the computational volume follow the luminosity
function given by equation~(40). Then, we compute the galaxy
parameters using equations~(48)--(50). 

\subsection{Computing the Evolution of the Beam}

\subsubsection{Building up a Sequence of Lens Planes}

Since each plane represents a different region of the universe,
the large-scale structure inside each plane should be uncorrelated 
with the large-scale structure inside the neighboring planes. 
This is clearly a problem if the lens planes originate from
one single calculation, since they would then
represent the same large-scale
structure at various evolutionary stages.
To solve this problem, we perform five independent
calculations for each cosmological model, by using five different sets of
initial conditions. We then choose randomly which calculation will provide
each lens plane, making sure that two consecutive lens planes never come
from the same calculation. To eliminate correlations even more, we make
use of the periodic boundary conditions by giving to the galaxy 
and background matter distributions
in each lens plane a random shift. This is equivalent to choosing randomly
on each lens plane the location where
the beam will hit. We could eliminate correlations even
more by rotating and/or reflecting the galaxy distributions before
projecting them onto the lens planes, using the 48-fold symmetry
of the cubic computational volume, but we consider this to be an overkill.

\subsubsection{The Contribution of the Background Matter}

In this paper, we use the term ``background 
matter'' to refer to the total matter
in the universe, if the presence of galaxies is ignored. Hence, the
distribution of background 
matter in the universe at various redshifts is what the
P$^3$M code simulates. To compute the effect of the background matter on the
propagation of the beam, we solve equation~(13) numerically on a 
two-dimensional grid (a similar technique was used by Blandford et al. [1991]).
We compute the right-hand side of equation~(13) on a square grid,
using the location of the particles provided by the P$^3$M code, and invert
equation~(13) using a Fast Fourier Transform method which is
essentially the method that the P$^3$M algorithm itself uses for solving
the three-dimensional Poisson equation. The details 
of the calculation are given in Appendix~A.

For all simulations presented in this paper, the comoving
size $L_{\rm box}$ of the computational volume was $128\,\rm Mpc$, and
the grid used for solving equation~(13) was $128\times128$ in size. Therefore,
the grid spacing at redshift $z$ was $h=1\,{\rm Mpc}/(1+z)$. This has the
effect of smoothing out any density fluctuation in the 
background matter on
scales below $1\,{\rm Mpc}/(1+z)$. This is consistent with the assumption
made by Jaroszy\'nski (1991) that the actual 
background matter distribution in the
universe at that scale should be smooth. Of course, galaxies contain
dark matter halos with are presumably smaller than this, but these dark
matter halos are taken into account in the galaxy profiles given by
equation~(47). In this subsection, we are considering the smoother
component of the background matter that has not been accumulated into
galactic halos.

\subsubsection{The Contribution of the Galaxies}

The calculation of the background matter potential described above
is sufficient for computing the effect of distant matter on
the propagation of the beam. However, at distances less than a
few Mpc's, we cannot ignore the fact that matter has collapsed
to form galactic-size objects which are much smaller than the
resolution of the P$^3$M algorithm or the algorithm used
for solving equation~(13). This is why we added galaxies to the
simulations using the method described in \S3.3. Each lens plane
contains about 40000 galaxies, but only the ones located near
the beam can have a significant effect on its evolution. This
enables us to greatly reduce the computation time by only 
including nearby galaxies. 

We identify one particular ray in the beam as being
the ``central ray.'' Then, in each lens plane, we only compute the
contribution of the galaxies which are within a projected distance
$r_{\rm cutoff}$ of the central beam. In this paper, we chose
$r_{\rm cutoff}=4\,{\rm Mpc}/(1+z)$. Hence, the mean number of
galaxies per lens plane included in the calculation is 
$40000(\pi4^2/128^2)=123$. Of course, the actual number varies
over a wide range because galaxies are clustered.

The calculation of the background matter contribution
takes the {\it total} matter in the system into account. Therefore, for every
galaxy we add to the calculation, we must subtract something in order
to conserve mass. We make the assumption that each galaxy has formed
by accumulating matter that was originally distributed over a region
of comoving radius $r_{\rm hole}=1\,\rm Mpc$, which is of order the
present mean spacing between galaxies. This accumulated matter should
be removed from the calculation. This is done by putting on the top
of each galaxy a ``hole'' with radius $r=r_{\rm halo}$ and negative
density, so that the combined mass of the galaxy and hole is
zero. In this model, each galaxy has formed by accumulating matter for
a region of fixed size, the more massive galaxies simply accumulating
more matter from that region. This model is of course crude. One might
suggest that more massive galaxies accumulate matter from a larger region.
However, the relation between the galaxy mass and the size of that
region is unknown at present, due to our limited understanding of the
galaxy formation process. Furthermore, this relation most certainly
depends of the environment (whether the galaxy forms in isolation or in
a cluster). Until better models for galaxy formation are available, it is
a reasonable approximation 
to use a constant value for $r_{\rm hole}$. To eliminate
spurious edge effects, we do not use a hole with a flat 
negative density profile. Instead, we use a Gaussian density profile
with a FWHM equal to $r_{\rm hole}$. The calculation of the
potential for the galaxies and the holes are given in Appendix~B.

\section{THE EXPERIMENTS}

\subsection{The Models}

We consider three different cosmological models: an
Einstein-de~Sitter model with $\Omega_0=1$, $\lambda_0=0$,
an open model with $\Omega_0=0.2$, $\lambda_0=0$,
and a flat, low-density model with $\Omega_0=0.2$, $\lambda_0=0.8$,
where $\Omega_0$ and $\lambda_0$ are the present values of the
density parameter and cosmological constant, respectively.
We set the present value $H_0$ of the
Hubble constant equal to $50\,\rm km\,s^{-1}Mpc^{-1}$ to avoid conflict
between the models and the measurements of globular cluster ages.
With these parameters, the age of the universe $t_0$ is
13.0~Gyr, 16.6~Gyr, and 21.04~Gyr for the Einstein-de~Sitter, open,
and cosmological constant models, respectively.

In all cases, 
the comoving length of the computational volume is $L_{\rm box}=128\,{\rm Mpc}$
(present length units).
The total mass of the system is
$M_{\rm sys}=3H_0^2\Omega_0L_{\rm box}^3/8\pi G=1.455\times10^{17}
\Omega_0{\rm M}_\odot$.
We use $64^3=262,144$ equal mass particles. The mass per particle 
is therefore $M_{\rm part}=M_{\rm sys}/64^3=5.551\times10^{11}{\rm M}_\odot$ 
for the Einstein-de~Sitter model and $1.110\times10^{11}{\rm M}_\odot$ 
for the other two models.

For all simulations, we use the Cold Dark Matter (CDM) power spectrum of
Bardeen et al. (1986), with the normalization of
Bunn, Scott, \& White (1995).
As mentioned in MPM, we use the same power spectrum for all three models, 
which is inconsistent, since the CDM power spectrum depends upon $\Omega_0$ 
and $\lambda_0$.
Our motivation for doing this is 
the following: Our goal in this paper
is not to find which model fits the observations of
the present universe better (we defer this to a forthcoming paper). 
Instead, we want to select cosmological models that 
will bracket the behavior of the large-scale structure formation process.
Using the same power spectrum for all models
allows us to investigate directly 
the effects of the growth rate and the age of the universe on the evolution of
the beam. In the same spirit, we are considering open models and 
models with a cosmological constant that are somewhat too extreme to 
agree with recent observations, which suggests that $\Omega_0$ is more likely
to be somewhere in the range 0.25--0.5 (Ostriker \&~Steinhardt 1995;
Martel, Shapiro, \&~Weinberg 1997, and references therein).
Models with a larger $\Omega_0$ and/or a smaller $\lambda_0$ would
reproduce observations better, but would resemble the Einstein-de~Sitter model
more than the ones we are considering, thus providing less insight on the 
effect of the cosmological parameters on the beam evolution.
The reader should therefore keep in mind that the power spectra we are 
using for the open and cosmological constant models are not consistent with 
a standard CDM model, and are chosen only for practical considerations. 

We ran 5 simulations for each of the three cosmological models, for
a total of 15 simulations. For each model, the 5~simulations differ
only in the ensemble of random phases used for generating
the initial conditions (see MPM). 
All simulations start at an initial redshift $z_i=24$, and end at $z=0$.
In all experiments, we propagate a beam composed of several light rays
backward in time, starting from the image plane, located near the observer,
and ending at the source plane, located at $z_S\simeq5$. There are 55 lens
planes for the Einstein-de~Sitter model, 73 for the open model, ant
96 for the cosmological constant model.

\subsection{First Experiment}

In this experiment, the light beams consist of 65 rays arranged in 
two concentric rings of 32 rays each,
plus a center ray. The rings' diameters on the image plane are
$2\times10^{-4}L_{\rm box}$ and $3\times10^{-4}L_{\rm box}$, corresponding
to angular sizes 1.54 and 2.30 arc seconds, respectively. 
We performed 500 calculations for each of the 3 cosmological models.
For each calculation, we used a different seed for the random number generator
that computes the random shifts of the lens planes. Hence, the calculations
within each model differ from one another in the location on
the lens planes where the beam hits.

Figure~3 shows the configuration of the beam on the source plane, located at 
$z_S\simeq5$,\footnote{The source planes were defined as being coincident
with the {\it next} lens plane, had we decided to propagate
the beam to higher redshifts. Hence $z_S$ was determined by using 
equation~(34) with $z'=z_S$, and $z=z_N$ being the redshift of the
last lens plane. The actual redshifts are $z_S=5.32$, $z_S=5.24$, 
and $z_S=5.08$,
for the Einstein-de~Sitter, open, and cosmological constant models, 
respectively}
for a subset of 27~calculations, 9 for each cosmological models.
The labels ``EdS,'' ``O,'' and ``$\Lambda$'' in this Figure and
all the subsequent ones identify the Einstein-de~Sitter model, the
open model, and the cosmological constant model, respectively. The
panels labeled ``NULL'' show for comparison
the configuration of the beam in the absence
of lensing, computed using equation~(2) with $\alphabf=0$.
The size of the panels is $4\,{\rm Mpc}/(1+z_S)\approx0.7\,{\rm Mpc}$.
The beam has a smaller diameter for the EdS model than the other two,
because of the dependence of the lensing equation upon the angular diameter 
distance. The deformation of the beam is comparable for the EdS
and $\Lambda$ models. This results from the combination of
two different effects that partly cancel each other: On one hand, the
large-scale structure is more developed in the EdS model than in the
$\Lambda$ model. On the other
hand, there are almost twice as many
lens planes between $z=0$ and $z=5$ in the $\Lambda$ model
than in the EdS model. 

Figure 4 shows the 
individual contribution of each lens plane to the shear 
as a function of the lens-plane redshift $z$
for 3 particular
runs, one for each model. 
These results were obtained by averaging the magnification matrix
over all 65 rays in the beam.
The sharp variations result from the absence of
correlation between neighboring
lens planes. The beam may experience a strong shear
in one particular plane simply because the beam happens to pass near a
large cluster. In order to eliminate this source of noise, we
average the shear over all 500 calculations for each model. The results
are shown in Figure~5. The lens planes that contribute most to the shear 
are located at intermediate redshifts, of order $z=1-2$,
for all three models. The contribution of lens planes located near the 
source or near the observer is significantly smaller.

Figure~6, shows the individual
contribution of each lens plane to
the magnification, as a function of the lens-plane redshift $z$
for 3 particular calculations, one for each model, using again
the average magnification matrix.
As in
Figure~4, the large fluctuations are caused by the absence of correlations
between consecutive lens planes. Figure~7 shows the result of averaging
the magnification over all 500 calculations for each model.
As for the shear, the lens planes that contribute the most to
the magnification are located at intermediate redshifts. 
Notice that the average magnification is almost always larger than unity.
This is not a violation of flux conservation. If we were averaging the
magnification matrix over {\it sources}, the average magnification would be
exactly unity. But we are instead averaging over {\it light rays}.
This clearly weights in favor of sources with $\mu>1$, since more
light rays originate from these sources, which is why they are magnified in
the first place. Hence, averaging the magnification matrix over light rays
instead of sources will gives $\langle\mu\rangle>1$.

The shear is much larger for the Einstein-de~Sitter model
than the other two models. Even though the large-scale structure is more
evolved for the Einstein-de~Sitter model, this cannot account for the 
difference. In particular, the cosmological constant model resembles
the Einstein-de~Sitter model much more than the open model at late
time (MPM). Hence, the differences between the structure in the
various models is not sufficient to explain the results shown on
Figure~5. The correct explanation is quite simple: though
all models contain the same mass {\it in galaxies}, the 
{\it total mass} is 5 times larger for the Einstein-de~Sitter model
than for the other two models, for which $\Omega_0=0.2$. We will
prove this affirmation below. Notice that even though the contribution
of individual lens planes is larger for the Einstein-de~Sitter model,
the number of such planes between $z=0$ and $z=5$ is smaller, so we do
not necessarily expect the cumulative shear and magnification for
distant sources to be larger for this model.

For all 1500 calculations, and all lens planes, we computed
the ratio $S_i^2/(1-\kappa_i)^2$, which measures the relative contributions
of the shear and convergence to the magnification (see eq.~[23]).
The largest value was $2.8\times10^{-5}$, implying that
the contribution of the shear to the magnification is totally negligible,
for all models, all calculations, and at all redshifts. This is
a well-known result (Lee \& Paczy\'nski 1990; Jaroszy\'nski et al. 1990).

The most interesting result that comes out of these calculations is the
fact that the largest contribution to 
both shear and magnification comes from matter located at intermediate 
redshifts. Equation~(6) shows that the critical surface density is large
for lens planes located near the image plane ($D_i$ small), or near the
source plane ($D_{iS}$ small), resulting in a small deflection potential
at small and large redshifts, and therefore ``favoring'' the lens planes
located at intermediate redshifts. However, large-scale structures grow
with time, an effect that favors lens planes located at small redshift. 
Furthermore, since we assume that the physical size of galaxies does not
evolve, the total cross section of the galaxies is larger in the past
since galaxies are closer to one another.
Hence, we expect a larger number of direct hits of galaxies by the beam
at larger redshift. On Figure~8, we plotted the number of galaxies hit by
the beam at each redshift, averaged over all 500 calculations for each
model (which explains why the numbers are not integer). Effectively, the
number of galaxies hit increases monotonically with redshift, with an average
of 1 galaxy hit by the beam at $z=5$ for the Einstein-de~Sitter model,
and 2 for the other models (the beams diverge more in the open and
cosmological models that in the Einstein-de~Sitter model [see Figs.~2 and~3],
so more galaxies get hit in these models).

This shows that, in spite of the fact that there are more structures
at small redshift and more galaxies hit at large redshift, the geometrical
factors in equation~(6) dominate, making the individual contribution
of lens planes to the shear and magnification larger at intermediate
redshift. Notice that on Figure~5, the average shear peaks at a 
redshift $z=1$ for the Einstein-de~Sitter model, while it peaks at
redshift of order $z=1.5$ for the other models. We interpret this result
as follows: in model with $\Omega_0<1$, the linear growth of the
density perturbation ``freezes out'' at redshift $z\sim\Omega_0^{-1}-1$,
whereas in an $\Omega_0=1$ model, linear growth persists all the way to the
present. Hence, in the Einstein-de~Sitter model, there is significant growth
taking place at small redshift, giving a ``boost'' to the shear
for lens planes located at $z=1$ relative to the ones located near $z=1.5-2$.
this effect results in a shift of the peak toward smaller redshift
in Figure~5. 

Figures~4--7 show the individual effect of each lens plane on the propagation
of the beam. To get the cumulative effect of all the lens planes, we
need to combine them using the formalism described in \S2.
Figure~9 (solid lines) shows the distributions of cumulative aspect ratios
computed using equation~(32). These distributions are very 
different for the different models.
The distribution is narrow for the open model and broad for
the other two models.
The distributions have a very similar shape for the Einstein-de~Sitter
and cosmological constant model, but the mean value of the distribution 
is significantly larger for the cosmological constant model.
Figure~10 (solid lines) shows the distributions of cumulative magnifications,
computed using equation~(33).
As for the distributions of aspect ratios, the distributions of
magnifications are broad and similar in shape for the Einstein-de~Sitter
and cosmological constant model, the latter one being shifted to larger values,
while the distribution is narrow for the open model.
These distributions are characterized by a sharp increase
on the low side and a more extended tail on the high side. They
are qualitatively in agreement with the analytical and numerical
results obtained by various authors (Schneider \& Weiss 1988a;
Lee \& Paczy\'nski 1990)

We ran an additional 450 calculations, 150 for each of the three 
cosmological model, with the source plane located at redshift $z_S=3$
instead of~5. 
Figures~9 and~10 (dotted lines) 
show the distributions of aspect ratios and magnifications 
for these calculations
(we multiplied the counts in each bin by $500/150=3.333$ to allow
a direct comparison). They are qualitatively very similar to the
distributions for sources located at $z_S=5$. The aspects ratios are smaller,
and the magnifications are closer to unity, but the relative similarity
and differences between the various models are the same. The only difference
is in the aspect ratios, where the distribution for the cosmological constant
model is not shifted to larger values compared with the Einstein-de~Sitter 
model. The distributions of aspect ratios peak at values of order 1.1, which
is somewhat large compared with observation of lensed quasars at that redshift.
This simply indicates that the standard CDM model normalized to COBE does
not reproduce observations well.

Returning to the 1500 calculations with $z_S=5$, we plot
in Figure~11 the average shear versus the
redshift of the lens plane, where
the shear is computed by
including either the contribution of the background matter only (solid curves)
or the contribution of the galaxies only (dashed curves). Notice that
the total shear, shown in Figure~5, is not equal to the sum of these
components, since it is the matrix elements $S_{11}$ and $S_{12}$,
and not the shear $S$, that add up. 
This figure shows that the contributions
of the background matter and galaxies are nearly identical for the 
open and cosmological constant model,
while the background matter contribution greatly exceeds the 
galaxies' contribution
for the Einstein-de~Sitter model. This shows the importance
of including the contribution of distant background
matter to the evolution of the beam,
something that was overlooked in some previous studies
(Schneider \& Weiss 1988a, b; Jaroszy\'nski 1991, 1992). We should point out,
however, than these results are obtained by averaging over 500 calculations
for each model. In one individual calculation, the effect of a single
massive galaxy might dominate over the effect of the background if the
beam hits or nearly hits that galaxy.

\subsection{Second Experiment}

In this experiment, the beam consists of $31^2$ light rays arranged in a 
square lattice. The spacing between rays on the source plane is 1 arcsecond, 
about the size of an extended radio source. 
We ran 10 simulations for each cosmological model. 
Figure~12 shows the final configuration of the beam, for 6
particular runs, two for each model. The overall deformation of the
square array into an irregular, 4-side polygon is caused by the background
matter. This deformation is large for the Einstein-de~Sitter
and cosmological constant models, and small for the open model,
consistent with the results shown in Figure~3.
In all cases, there are small regions
on the source plane where several rays converge. 
This convergence results from the presence of galaxies.
A source located in one
of these regions has 
in general multiple images. We could use these results to estimate
the fraction of high-redshift quasars with multiple images simply by computing
the fraction of the source plane which is covered by these regions. 
{From} Figure~12, we can estimate that a few percent of high-redshift 
quasars have multiple images, by counting the number of rays in the
regions of convergence.\footnote{Of course, we cannot compare
this prediction with observations, since too few quasars at $z\sim5$ are
known.}
To get statistically significant results, however, we need to perform many
more calculations. Also, in order to estimate the multiplicity (double,
triple, ...) of each lensing event, we need to improve the resolution
of these calculations, by increasing the number of rays per unit solid
angle in the beam. Hence, the results we are presenting in this subsection are
for illustrative purpose only.
A detailed study of the statistics of multiple images of quasars
will be presented in a forthcoming paper.

We redid the calculation shown in the
middle left panel of Figure~12 with twice the resolution per dimension
($63\times63$~rays). The results are shown in Figure~13. Increasing the
resolution reveals several additional regions of convergence. Only two such 
regions are clearly visible in Figure~12, whereas there are 6 clearly
visible regions of convergence in Figure~13.

\section{SUMMARY AND CONCLUSION}

We have developed a numerical algorithm for studying the propagation of
light in inhomogeneous universes. 
This is the first algorithm that combines fully nonlinear N-body simulations
of large-scale structure formation with realistic distributions of galaxies,
reproduces the 2-point correlation function of galaxies, and takes
the morphology-density relation into account when
ascribing morphological types to galaxies. As a result, this algorithm
reproduces the matter distribution in the universe with a level
of realism that is unprecedented. The density structure
of the lens planes is simulated over 9 orders of magnitude in
length, from the size of superclusters and voids down to the core
radii of small galaxies. 

We use this new algorithm to study the propagation of light
beams in inhomogeneous universes, 
for three different cosmological models: the Einstein-de~Sitter model, an
open model with $\Omega_0=0.2$, and a flat cosmological constant model
with $\Omega_0=0.2$, $\lambda_0=0.8$. We performed 1981 simulations, 
propagating light beams back in time, up to a redshift of $z=3$ or~5.

The average magnitude of shear and magnification amongst models are shown to 
be different, with the values for the  
cosmological constant model being significantly
larger than for the other two models, for
sources located at $z=5$. The 
contribution of individual lens planes to
the shear and magnification is larger for planes located at
intermediate redshift, of order $1-2$, even though structures
are more evolved at low redshift and direct hits of galaxies are
more frequent at high redshift.
The lens planes providing the largest average contribution
to the shear are located at
lower redshift for the Einstein-de~Sitter model than for the other two models.
The contribution of distant background matter to the shear
is as important as the contribution of nearby galaxies (see Fig.~11)
for low~$\Omega_0$ models, and significantly more important for
the Einstein-de~Sitter model.
These results, combined with observations of lensed quasars, might
eventually put limits on the value of the cosmological parameters
$\Omega_0$ and $\lambda_0$. 

This paper has focussed on the description of the method.
Applications of this method to various problems will be presented
in forthcoming papers. These include studying the
statistics of multiple imaging of quasars, and their dependence upon
the source redshift (Premadi, Martel, \& Matzner 1997a),
performing a cosmological parameter survey, for several values of
the cosmological parameters $H_0$, $\Omega_0$, and $\lambda_0$,
and for various density perturbation spectra (Premadi, Martel, \&~Matzner
1997b), and modifying the algorithm to include the effect of
microlensing by stars inside galaxies (Martel, Premadi, \& Matzner 1997b)

\acknowledgments
 
This work benefited from stimulating discussions with
Alan Dressler, Daniel Holz, and Inger J\o rgensen.
We are pleased to acknowledge the support of NASA Grant NAG5-2785,
NSF Grants PHY93~10083 and ASC~9504046,
the University of Texas High Performance Computing Facility
through the office of the vice president for research,
and Cray Research.

\appendix
\section{THE POTENTIAL OF THE BACKGROUND MATTER}

We compute the potential of the background matter by solving equation~(13)
numerically. We first rewrite this equation as
\begin{equation} 
\nabla^2\psi={2(\sigma-\bar\sigma)\over\sigma_{\rm crit}}\equiv 2Q\,,
\end{equation}

\noindent where we have introduced the mean surface density $\bar\sigma$
to be consistent with the filled-beam approximation,
which requires that the mean surface density
in each lens plane vanishes (Schneider \& Weiss 1988a, b). 

To solve this equation,
we first use the location of the particles provided by
the P$^3$M code to compute the source
term $Q$ in equation~(A1) on a square grid of size $N\times N$, using
the {\it Triangular Shaped Cloud} (TSC) assignment scheme (Hockney
\&~Eastwood 1981). The values $Q_{k,l}$ of $Q$ at each grid point $(k,l)$ is 
given by
\begin{equation}
Q_{k,l}=-{\bar\sigma\over\sigma_{\rm crit}}+{m\over h^2\sigma_{\rm crit}}
\sum_p W\big(|x_p-x_{k,l}|\big)W\big(|y_p-y_{k,l}|\big)\,,
\qquad k,l = 0,1,\ldots,N-1\,,
\end{equation}

\noindent where $m$ is the particle mass, 
$h\equiv L_{\rm box}/N$ is the grid spacing, 
$x_p$, $y_p$ are the coordinates of particle~$p$, and 
$x_{k,l}\equiv(k+1/2)h$, $y_{k,l}\equiv(l+1/2)h$ are the coordinates of
the grid point $(k,l)$. For the TSC assignment scheme, the weight
function $W$ is given by
\begin{equation}
W(s)=\cases{
{3\over4}-s^2\,,&$s\leq{1\over2}$\,;\cr
{1\over2}({3\over2}-s)^2\,,&${1\over2}\leq s\leq{3\over2}$\,;\cr
0\,,&$s>{3\over2}$\,.\cr}
\end{equation}

Once the function $Q$ has been computed on the grid, we solve equation~(A1)
using a finite difference method. The finite-difference form of equation~(A1)
is
\begin{equation}
\psi_{k-1,l}+\psi_{k+1,l}+\psi_{k,l-1}+\psi_{k,l+1}-4\psi_{k,l}=2Q_{k,l}\,,
\end{equation}

\noindent where we used the standard 5-point formula for the 
two-dimensional Laplacian.
Since the grid has periodic boundary conditions, we can easily invert
this equation using Fourier techniques. We write the potential as
\begin{equation}
\psi_{k,l}={1\over N^2}\sum_{m=0}^{N-1}\sum_{n=0}^{N-1}
\hat\psi_{m,n}e^{-2\pi i(km+ln)/N}\,,
\end{equation}

\noindent where $\hat\psi$ is the discrete Fourier transform of $\psi$
(not to be confused with the dimensional potential).
We use a similar expression for the source term $Q$. We eliminate
$\psi$ and $Q$ in equation~(A4) and get, after some algebra,
\begin{equation}
\hat\psi_{m,n}=-{\hat Q_{m,n}\over2
\biggl(\displaystyle\sin^2{\pi m\over N}+\sin^2{\pi n\over N}\biggr)}\,.
\end{equation}

\noindent To solve for the potential, we compute $\hat Q$ by taking
the two-dimensional Fast Fourier Transform (FFT) of $Q$ on the grid. We
then use equation~(A6) to compute the function $\hat\psi$ on the grid,
and take the two-dimensional inverse FFT on the grid to
get $\psi$. Once we know the potential on the grid, we can 
compute its first and second derivatives using standard finite
difference formulae,
\begin{eqnarray}
\biggl({\partial\psi\over\partial x}\biggr)_{k,l}&=&
{\psi_{k+1,l}-\psi_{k-1,l}\over2h}\,,\\
\biggl({\partial\psi\over\partial y}\biggr)_{k,l}&=&
{\psi_{k,l+1}-\psi_{k,l-1}\over2h}\,,\\
\biggl({\partial^2\psi\over\partial x^2}\biggr)_{k,l}&=&
{\psi_{k-1,l}+\psi_{k+1,l}-2\psi_{k,l}\over h^2}\,,\\
\biggl({\partial^2\psi\over\partial y^2}\biggr)_{k,l}&=&
{\psi_{k,l-1}+\psi_{k,l+1}-2\psi_{k,l}\over h^2}\,,\\
\biggl({\partial^2\psi\over\partial x\partial y}\biggr)_{k,l}&=&
{\psi_{k-1,l-1}+\psi_{k+1,l+1}-\psi_{k+1,l-1}-\psi_{k-1,l+1}\over4h^2}\,.
\end{eqnarray}

\noindent This gives us the derivatives at the grid points. To interpolate
these derivatives at the location of the rays, we use again the TSC
assignment scheme,
\begin{equation}
f(x,y)=\sum_{k,l}W\big(|x-x_{k,l}|\big)W\big(|y-y_{k,l}|\big)f_{k,l}\,,
\end{equation}

\noindent where $x,y$ are now the coordinates of the ray, the sum is
on the grid points, and $f$ stands for any of the derivatives given
by equations~(A7)--(A11).

\section{THE POTENTIAL OF THE GALAXIES}

The deflection potential caused by a galaxy is given by equation~(5),
\begin{equation}
\hat\psi_{\rm gal}(\xibf)={4G\over c^2}\int\!\!\!\int\sigma(\xibf')
    \ln|\xibf-\xibf'|d^2\xi'\,.
\end{equation}

\noindent
We set ${\bf r}=\xibf-\xibf_{\rm gal}$ and ${\bf s}=\xibf'-\xibf_{\rm gal}$, 
where $\xibf_{\rm gal}$ is the position vector of the center of the
galaxy on the lens plane. We get
\begin{equation}
\hat\psi_{\rm gal}({\bf r})={4G\over c^2}\int\!\!\!\int\sigma({\bf s})
    \ln|{\bf r}-{\bf s}|d^2s\,.
\end{equation}

\noindent We write the argument of the logarithm as
\begin{equation}
|{\bf r}-{\bf s}|=(r^2+s^2-2rs\cos\theta)^{1/2}\,,
\end{equation}

\noindent
with $\theta$ being the angle between the two vectors ${\bf r}$ 
and ${\bf s}$. Equation~(B2) then becomes
\begin{eqnarray}
\hat\psi_{\rm gal}({\bf r})&=&{4G\over c^2}\int_0^{2\pi}d\theta
               \int_0^\infty\sigma({\bf s})s
               \biggl[{1\over2}\ln(r^2+s^2-2rs\cos\theta)\biggr]ds\nonumber\\
&=&{4\pi G\over c^2}
\int_0^{r_{\max}}\sigma(s)s\ln\left({r^2+s^2+|r^2-s^2|\over2}\right)ds
\,,
\end{eqnarray}

\noindent
where we used the fact the $\sigma({\bf s})$ is a function of
$s=|{\bf s}|$ only to perform the integration over~$\theta$. If the point 
$\xibf$ is located outside the density distribution, then $r>s$ for all 
$s\leq r_{\max}$,
and equation~(B4) reduces to
\begin{equation}
\hat\psi_{\rm gal}({\bf r})={8\pi G\ln r\over c^2}
\int_0^{r_{\max}}\sigma(s)s\,ds\,.
\end{equation}

\noindent If instead the point $\xibf$ is 
interior to the density distribution,
we must divide the integration interval in equation~(B4) into two parts,
\begin{equation}
\hat\psi_{\rm gal}({\bf r})={8\pi G\ln r\over c^2}\int_0^r\sigma(s)s\,ds
+{8\pi G\over c^2}\int_r^{r_{\max}}\sigma(s)s\ln s\,ds
\,.
\end{equation}

\noindent
The surface density of an isothermal sphere is given by
\begin{equation}
\sigma(s)=\cases{\displaystyle
{v^2\over4G(s^2+r_c^2)^{1/2}}\,,&$s\leq r_{\max}$;\cr
0\,,&$s>r_{\max}$;\cr}
\end{equation}

\noindent 
(eq.~[47]). We substitute this expression in equations~(B5), and~(B6), and
integrate. For the case $r>r_{\max}$ (eq.~[B5]), we get
\begin{equation}
\hat\psi_{\rm gal}({\bf r})
=2\pi u^2\ln r\Bigl[(r_{\max}^2+r_c^2)^{1/2}-r_c\Bigr]\,,
\end{equation}

\noindent where $u\equiv v/c$.
The first and second derivatives are then given by
\begin{eqnarray}
{\partial\hat\psi_{\rm gal}\over\partial x}&=&{2\pi u^2x\over r^2}
\Bigl[(r_{\max}^2+r_c^2)^{1/2}-r_c\Bigr]\,,\\
{\partial\hat\psi_{\rm gal}\over\partial y}&=&{2\pi u^2y\over r^2}
\Bigl[(r_{\max}^2+r_c^2)^{1/2}-r_c\Bigr]\,,\\
{\partial^2\hat\psi_{\rm gal}\over\partial x^2}&=&{2\pi u^2(y^2-x^2)\over r^4}
\Bigl[(r_{\max}^2+r_c^2)^{1/2}-r_c\Bigr]\,,\\
{\partial^2\hat\psi_{\rm gal}\over\partial x\partial y}&=&
-{4\pi u^2xy\over r^4}\Bigl[(r_{\max}^2+r_c^2)^{1/2}-r_c\Bigr]\,,\\
{\partial^2\hat\psi_{\rm gal}\over\partial y^2}&=&{2\pi u^2(x^2-y^2)\over r^4}
\Bigl[(r_{\max}^2+r_c^2)^{1/2}-r_c\Bigr]\,,
\end{eqnarray}

\noindent where $x$ and $y$ are the components of ${\bf r}$.
For the case $r<r_{\max}$ (eq.~[B6]), we get, after some algebra
\begin{eqnarray}
\hat\psi_{\rm gal}&=&2\pi u^2\Biggl\{
(\ln r_{\max}-1)(r_{\max}^2+r_c^2)^{1/2}
+r_c\ln\biggl[{r_c+(r_{\max}^2+r_c^2)^{1/2}\over r_{\max}}\biggr]\nonumber\\
&+&(r^2+r_c^2)^{1/2}-r_c\ln\Bigl[r_c+(r^2+r_c^2)^{1/2}\Bigr]\Biggr\}
\,.
\end{eqnarray}

\noindent Notice that the first two terms are constant and thus
do not contribute to the derivatives.
The first and second derivatives are given by
\begin{eqnarray}
{\partial\hat\psi_{\rm gal}\over\partial x}&=&{2\pi u^2x\over r^2}
\Bigl[(r^2+r_c^2)^{1/2}-r_c\Bigr]\,,\\
{\partial\hat\psi_{\rm gal}\over\partial y}&=&{2\pi u^2y\over r^2}
\Bigl[(r^2+r_c^2)^{1/2}-r_c\Bigr]\,,\\
{\partial^2\hat\psi_{\rm gal}\over\partial x^2}&=&{2\pi u^2\over r^4}
\biggl\{(y^2-x^2)\Bigl[(r^2+r_c^2)^{1/2}-r_c\Bigr]
+{x^2r^2\over(r^2+r_c^2)^{1/2}}\biggr\}\,,\\
{\partial^2\hat\psi_{\rm gal}\over\partial x\partial y}&=&
-{4\pi u^2xy\over r^4}\biggl[(r^2+r_c^2)^{1/2}-r_c
-{r^2\over2(r^2+r_c^2)^{1/2}}\biggr]\,,\\
{\partial^2\hat\psi_{\rm gal}\over\partial y^2}&=&{2\pi u^2\over r^4}
\biggl\{(x^2-y^2)\Bigl[(r^2+r_c^2)^{1/2}-r_c\Bigr]
+{y^2r^2\over(r^2+r_c^2)^{1/2}}\biggr\}\,.
\end{eqnarray}

To conserve mass, we superpose on the top of each galaxy a ``hole'' of
negative density, which represents the matter that has been used up
to form the galaxy. This hole has a volume
density given by 
\begin{equation}
\rho_{\rm hole}(R)=-{M\over\pi^{3/2}r_{\rm hole}^3}e^{-R^2/r_{\rm hole}^2}\,,
\end{equation}

\noindent where $M$ is the mass of the galaxy, and $R$ is
the three-dimensional radial distance. The normalization constant
in equation~(B20) was chosen such that the total mass of the galaxy and
hole vanishes. The projected surface density is given by
\begin{equation}
\sigma_{\rm hole}(r)=\int_{-\infty}^\infty\rho_{\rm hole}(z)dz
=-{Me^{-r^2/r_{\rm hole}^2}\over\pi r_{\rm hole}^2}\,,
\end{equation}

\noindent where $z=(R^2-r^2)^{1/2}$.
We substitute this density profile in equation~(B2), and integrate.
The angular part of the integration is the same as for the galaxies.
After some algebra, we get
\begin{equation}
\psi_{\rm hole}=-{M(1-e^{-r^2/r_{\rm hole}^2})\ln r\over\pi}
-{2M\over\pi r_{\rm hole}^2}\int_r^\infty e^{-s^2/r_{\rm hole}^2}s\ln s\,ds\,.
\end{equation}

\noindent The last integral cannot be solved using elementary functions.
This is not a problem, since we are only interested in the derivatives
of the potential. After differentiation, we get
\begin{eqnarray}
{\partial\psi_{\rm hole}\over\partial x}&=&
-{M(1-e^{-w^2})x\over\pi r^2}\,,\\
{\partial\psi_{\rm hole}\over\partial y}&=&
-{M(1-e^{-w^2})y\over\pi r^2}\,,\\
{\partial^2\psi_{\rm hole}\over\partial x^2}&=&
-{M\over\pi r^4}\left[(y^2-x^2)(1-e^{-w^2})+2x^2w^2e^{-w^2}\right]\,,\\
{\partial^2\psi_{\rm hole}\over\partial y^2}&=&
-{M\over\pi r^4}\left[(x^2-y^2)(1-e^{-w^2})+2y^2w^2e^{-w^2}\right]\,,\\
{\partial^2\psi_{\rm hole}\over\partial x\partial y}&=&
{2Mxy\over\pi r^4}\left[1-e^{-w^2}-w^2e^{-w^2}\right]\,,
\end{eqnarray}

\noindent where $w\equiv r/r_{\rm halo}$. Notice that these derivatives
are well-behaved at $r=0$. By combining equations~(B9)--(B13) with
equation~(B23)--(B27), we see that the combined potential of the
galaxy and hole drops to zero as $r$ goes to infinity. 
In practice, we neglect the combined potential at distances larger than 
$3r_{\rm hole}$.

%

%

\clearpage
\begin{center}
Figure Captions
\end{center}

\figcaption{Schematic diagram illustrating the multiple-lens geometry, for
the particular case of two lens planes. The distances $D_1$, $D_2$, 
$D_S$, $D_{1S}$, and~$D_{2S}$  are angular diameter
distances. All angles are greatly exaggerated for clarity.}

\figcaption{Angular distance $D_S$ between the source and the observer,
versus redshift $z_S$ of the source, for the cosmological constant model
(top curve), the open model (middle curve), and the Einstein-de~Sitter
model (bottom curved). $R_0=c/H_0$ is the Hubble radius.}

\figcaption{Configuration of the beam on the source plane, for a subset
of the 1500 calculations described in \S4.2. The labels
``EdS,'' ``O,'' and ``$\Lambda$'' refer to the Einstein-de~Sitter model,
the open model, and the cosmological constant model, respectively.
The panels labeled ``NULL'' show the configuration the beam would
have in the absence of lensing.}

\figcaption{Shear versus lens redshift $z$, for three particular runs,
one for each model. Top panel: Einstein-de~Sitter model; middle panel: 
open model; bottom panel: cosmological constant model.}

\figcaption{Average shear versus lens redshift $z$, obtained by
averaging over all 500 runs
for each model. Top panel: Einstein-de~Sitter model; middle panel: 
open model; bottom panel: cosmological constant model.}

\figcaption{Magnification versus lens redshift $z$, for three particular runs,
one for each model. Top panel: Einstein-de~Sitter model; middle panel: 
open model; bottom panel: cosmological constant model.}

\figcaption{Average magnification versus lens redshift $z$, obtained by
averaging over all 500 runs
for each model. Top panel: Einstein-de~Sitter model; middle panel: 
open model; bottom panel: cosmological constant model.}

\figcaption{Mean number of galaxies hit by the circular beam versus
galaxy redshift $z$, obtained by averaging over all 500 runs for each model.
Top panel: Einstein-de~Sitter model; middle panel: 
open model; bottom panel: cosmological constant model.}

\figcaption{Distribution of image aspects ratios for circular
sources located at $z=5$ (solid lines) and $z=3$ (dotted lines).
Top panel: Einstein-de~Sitter model; middle panel: 
open model; bottom panel: cosmological constant model.
The counts for the $z=3$ models have been multiplied by 3.333 for
comparison.}

\figcaption{Distribution of image magnifications for circular
sources located at $z=5$ (solid lines) and $z=3$ (dotted lines).
Top panel: Einstein-de~Sitter model; middle panel: 
open model; bottom panel: cosmological constant model.
The counts for the $z=3$ models have been multiplied by 3.333 for
comparison.}

\figcaption{Average shear versus lens redshift $z$, obtained by
averaging over all 500 runs
for each model. Top panel: Einstein-de~Sitter model; middle panel: 
open model; bottom panel: cosmological constant model.
The solid curves show the results obtained by including only the
contribution of the background matter to the shear. The dashed curves show the
results obtained by including only the contributions 
of the galaxies to the shear}

\figcaption{Location of the rays on the source plane at $z=5$, 
for a subset of the calculations described in \S4.3. Top panels:
Einstein-de~Sitter model; middle panels: open model; bottom panels:
cosmological constant model. The middle and bottom panels are plotted on the
same scale. The top panels have been enlarged relative to the other
panels for clarity.}

\figcaption{Location of the rays on the source plane at $z=5$,
for the high resolution calculation described in \S4.3.
}

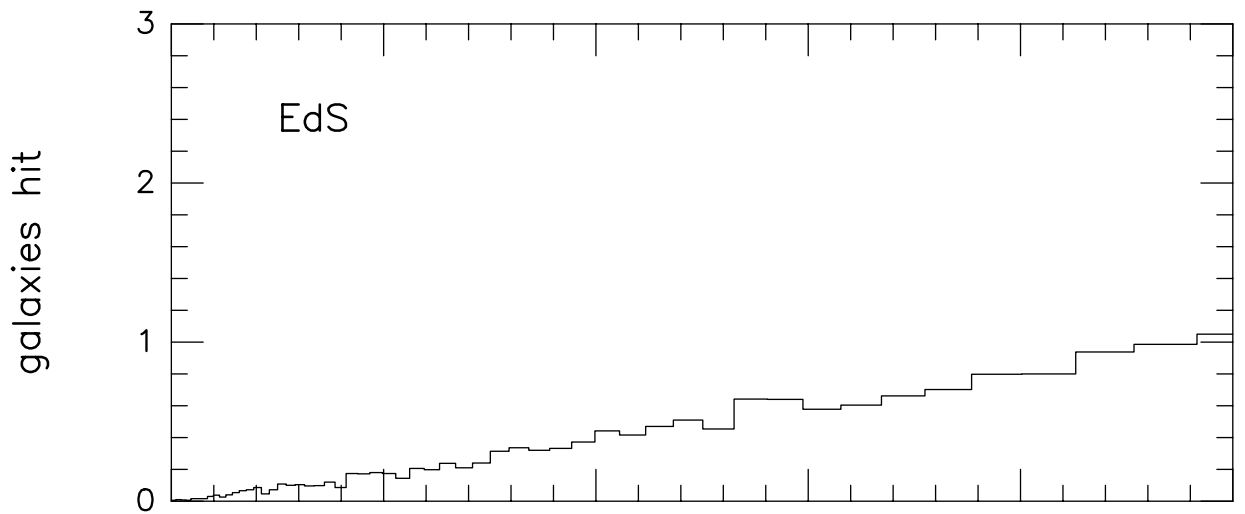
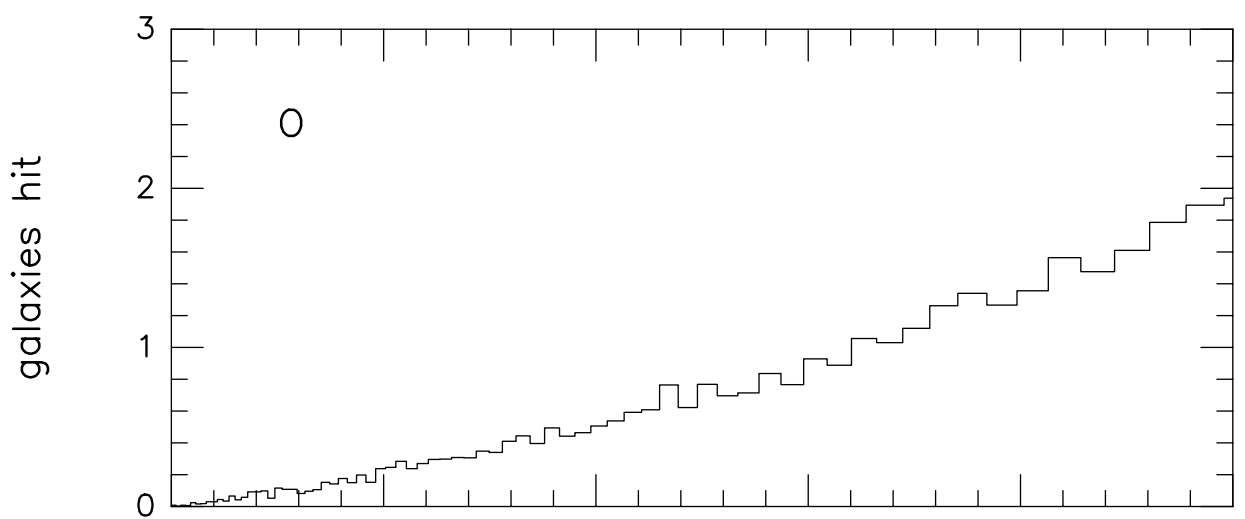
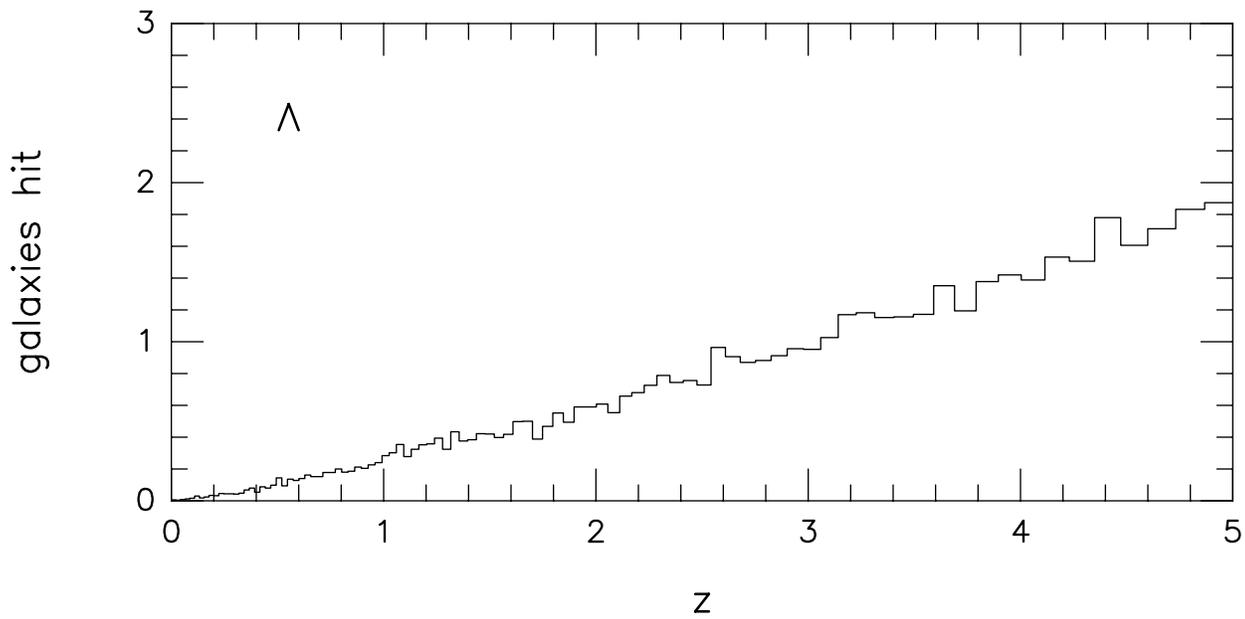
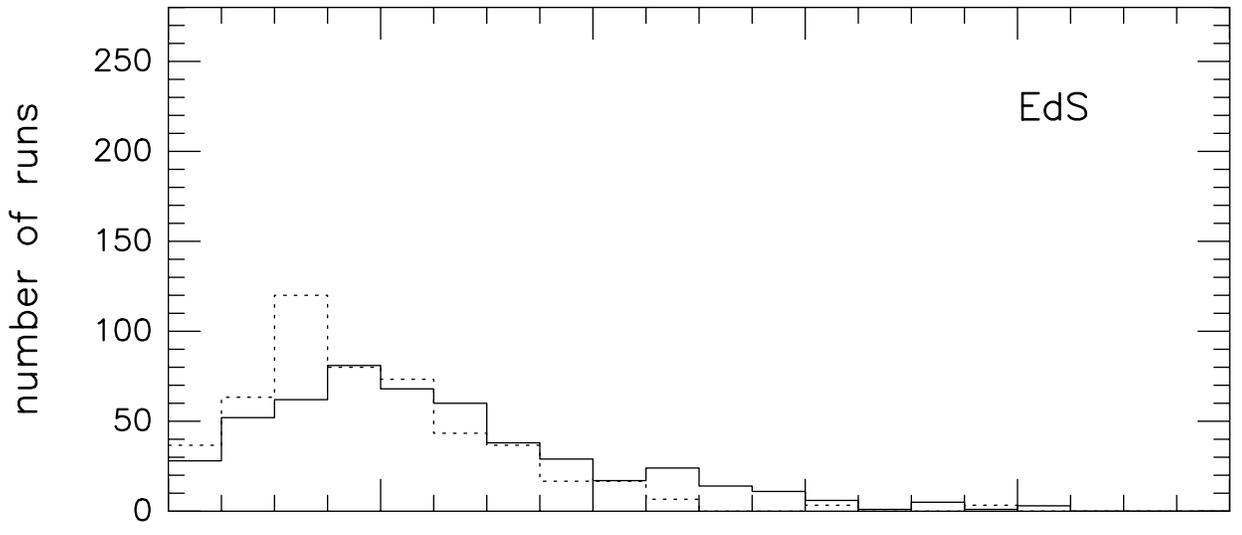
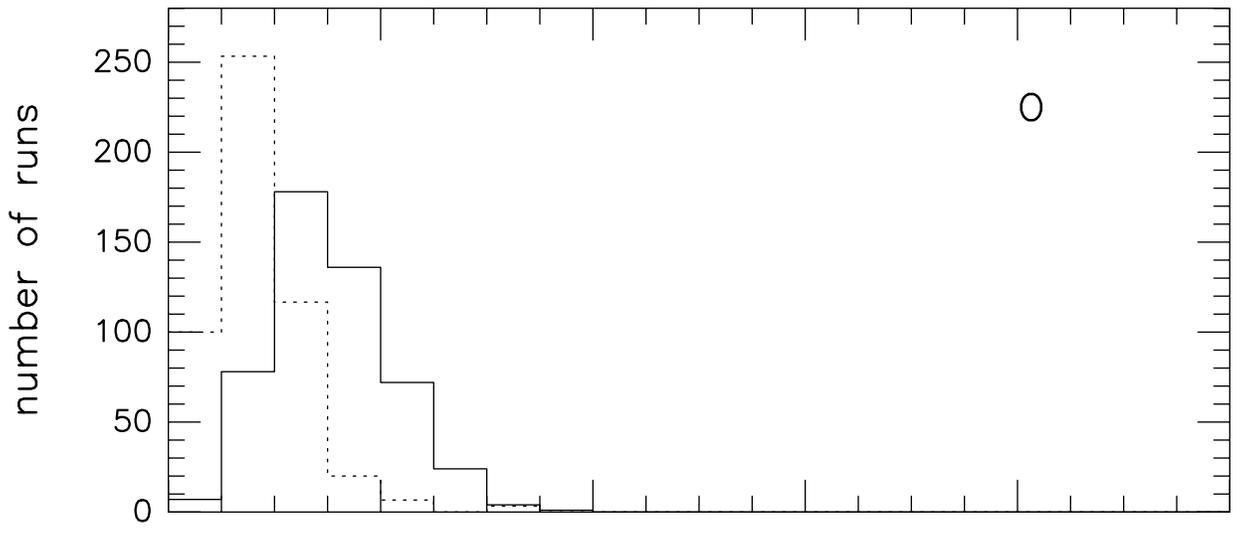
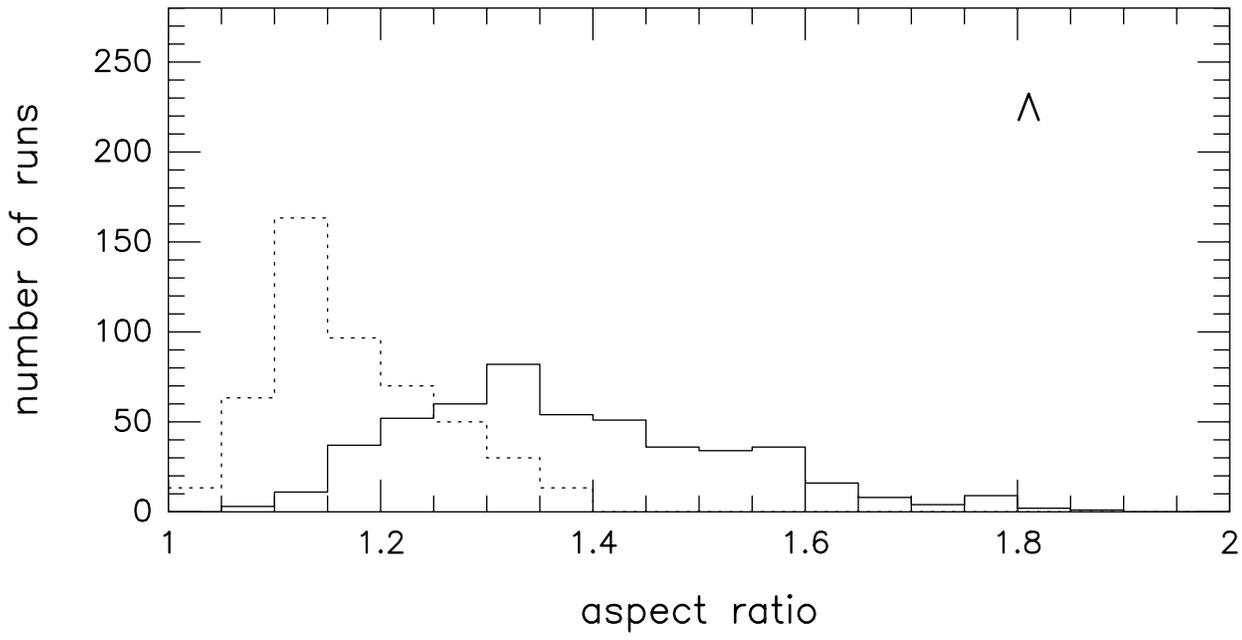
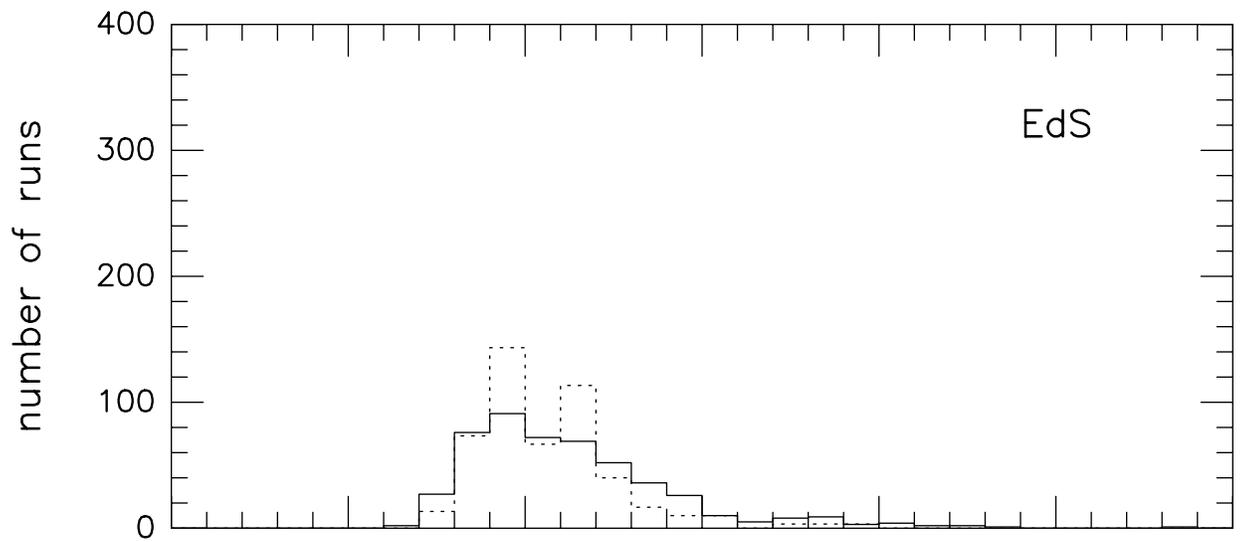
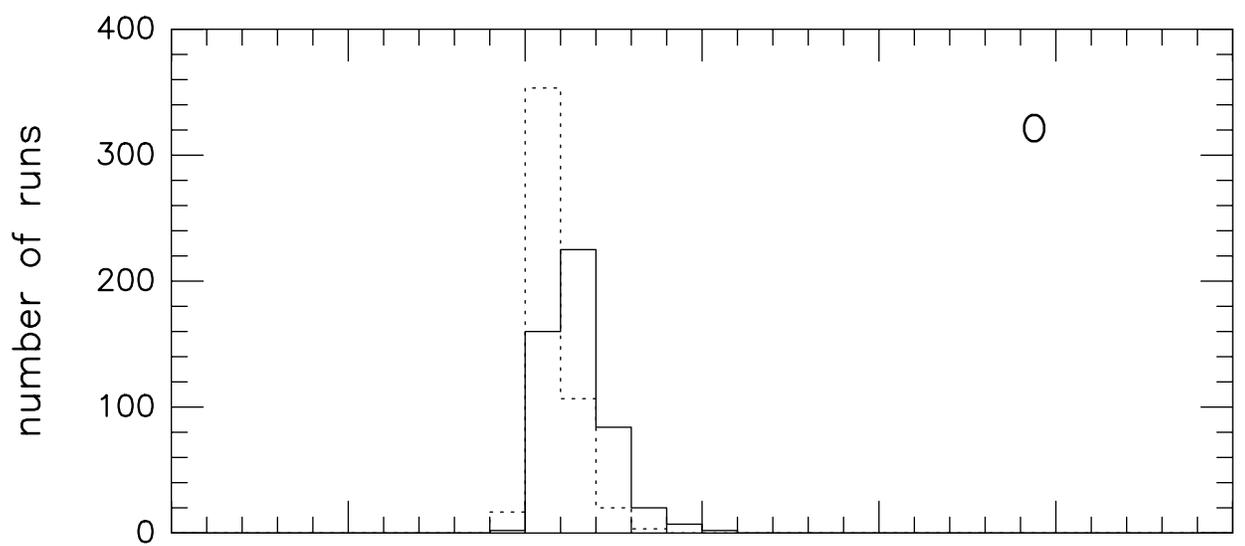
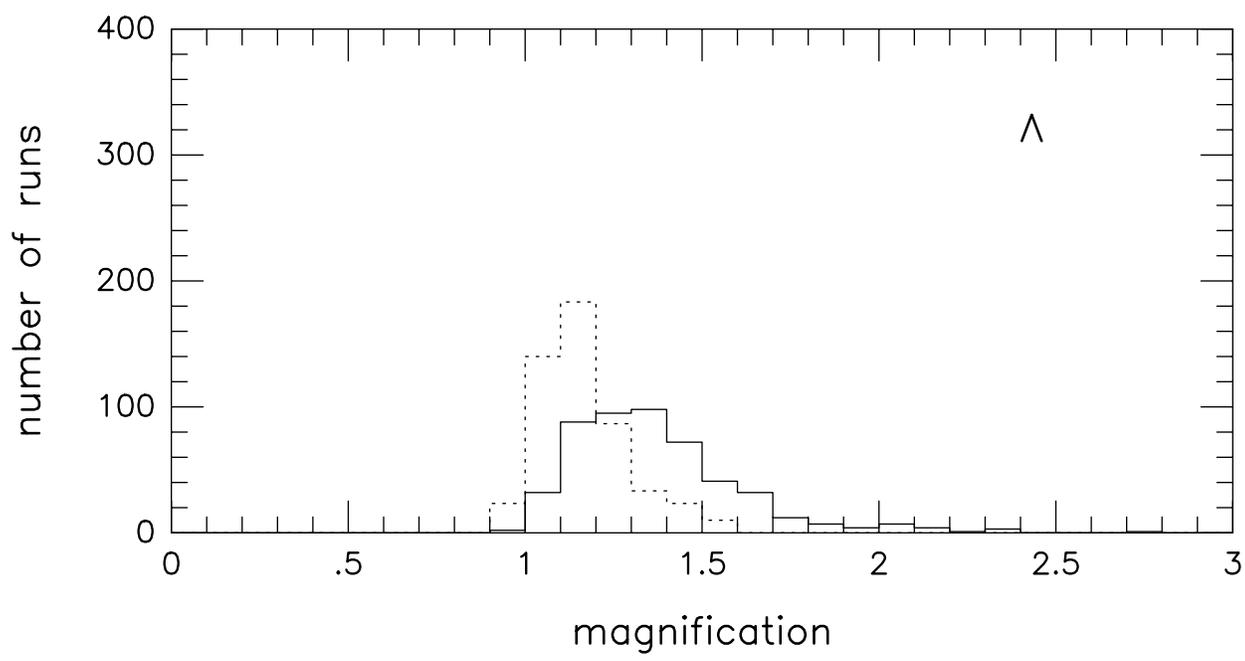
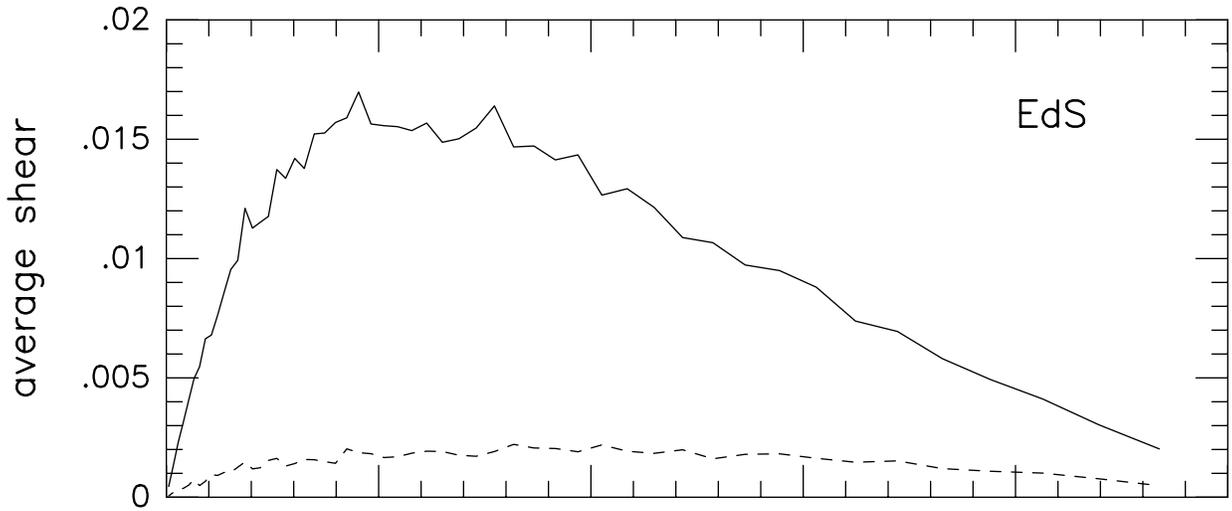
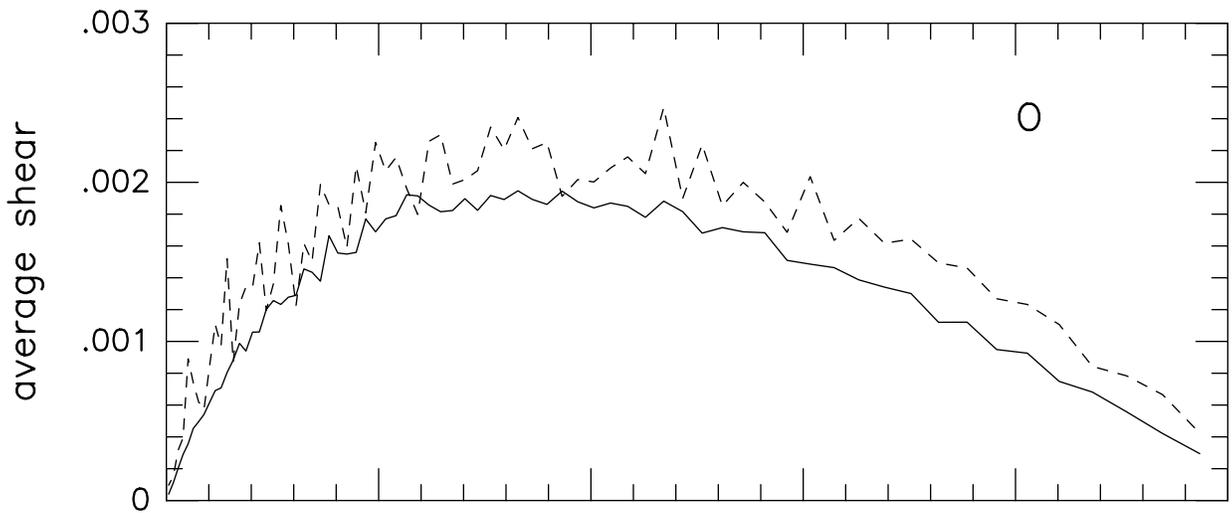
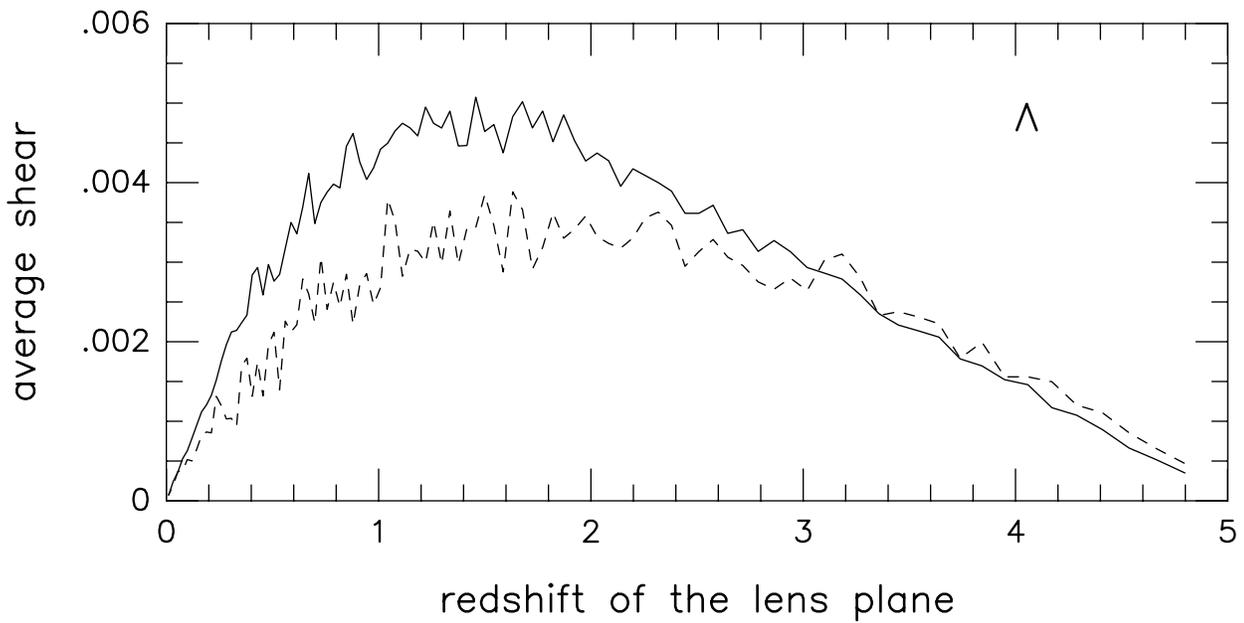
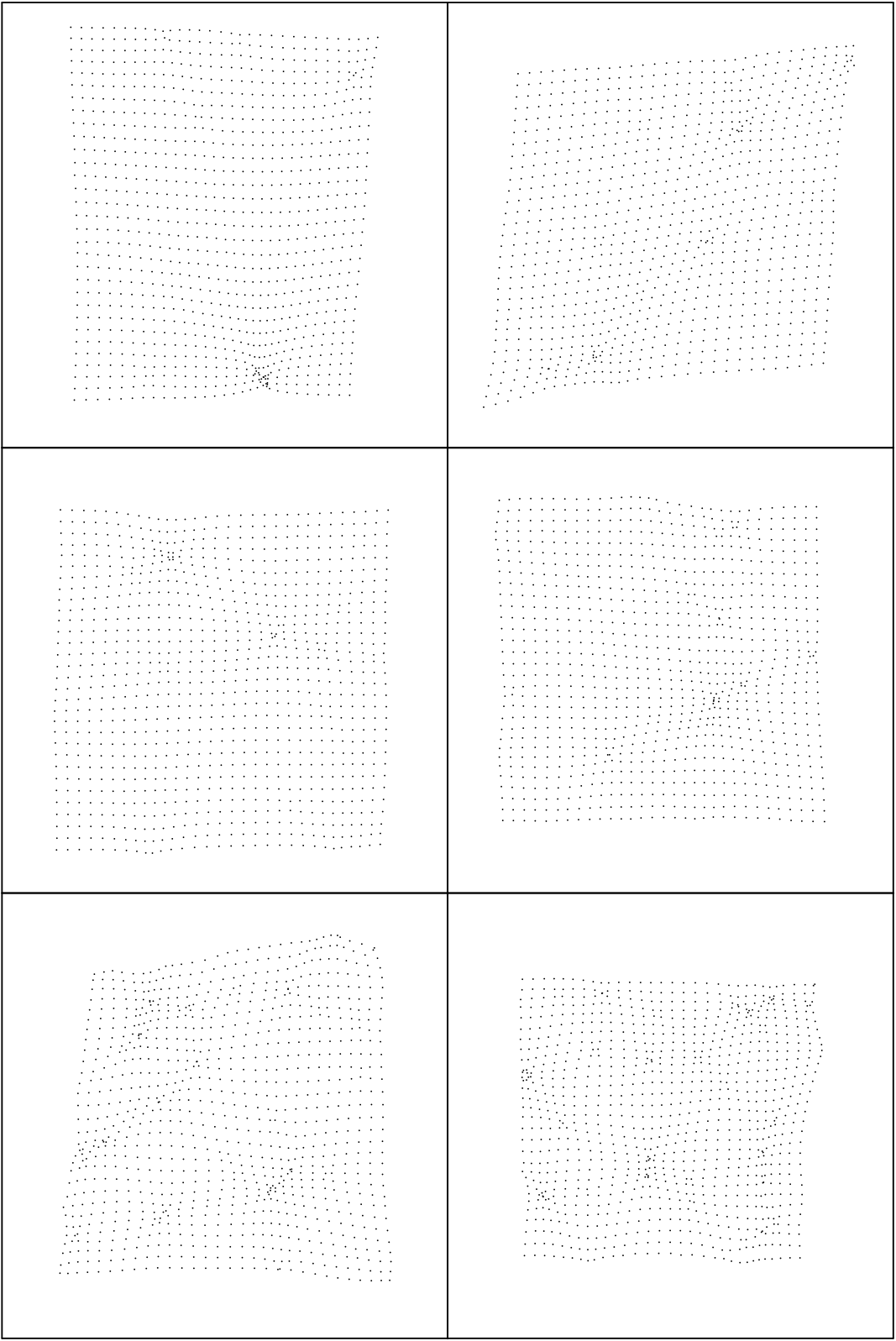
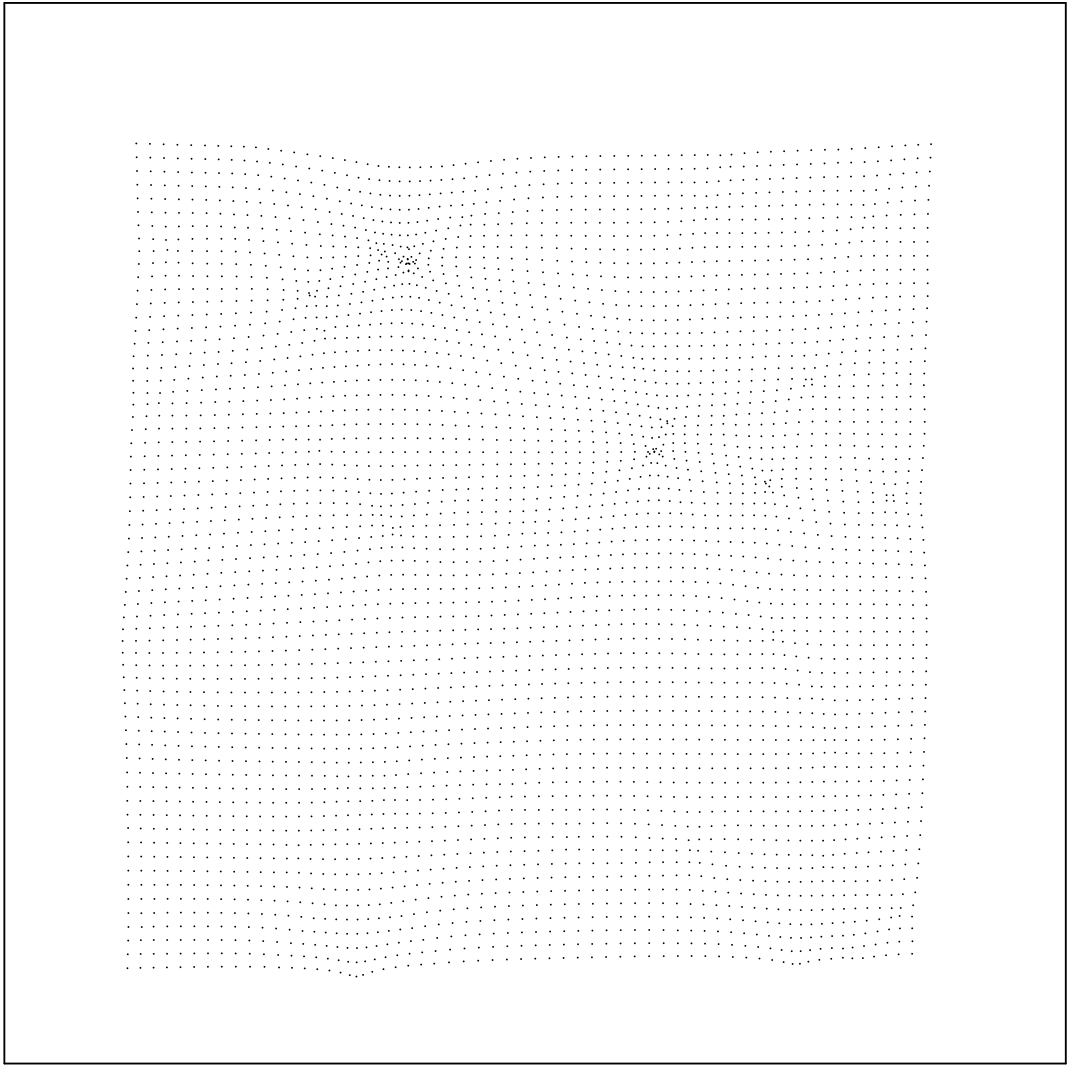
\end{document}